\documentclass[fleqn,usenatbib]{mnras}

\usepackage{newtxtext,newtxmath}

\usepackage[T1]{fontenc}

\DeclareRobustCommand{\VAN}[3]{#2}
\let\VANthebibliography\thebibliography
\def\thebibliography{\DeclareRobustCommand{\VAN}[3]{##3}\VANthebibliography}

\usepackage{graphicx}	
\usepackage{amsmath}
\usepackage{CJK}

\title[GPs from J1823$-$3021A with MeerKAT UHF]{
Detection of over 37,000 giant pulses per hour from PSR J1823$-$3021A with UHF baseband observations from MeerKAT}

\author[Ho et al.]{
Simon C.-C. Ho (何建璋),$^{1,2,3,4}$\thanks{E-mail: Simon.Ho@anu.edu.au}
Matthew Bailes,$^{2,3}$
Chris Flynn$^{2,3}$
and Federico Abbate$^{5,6}$
\\
$^{1}$Research School of Astronomy and Astrophysics, The Australian National University, Canberra, ACT 2611, Australia\\
$^{2}$Centre for Astrophysics and Supercomputing, Swinburne University of Technology, Hawthorn, VIC 3122, Australia\\
$^{3}$OzGrav: The Australian Research Council Centre of Excellence for Gravitational Wave Discovery, Hawthorn, VIC 3122, Australia\\
$^{4}$ASTRO3D: The Australian Research Council Centre of Excellence for All-sky Astrophysics in 3D, ACT 2611, Australia\\
$^{5}$INAF-Osservatorio Astronomico di Cagliari, Via della Scienza 5, I-09047 Selargius (CA), Italy\\
$^{6}$Max-Planck-Institut f{\"u}r Radioastronomie, Auf dem H{\"u}gel 69, D-53121, Bonn, Germany\\
}

\pubyear{2024}

\begin{document}
\begin{CJK*}{UTF8}{gbsn}
\label{firstpage}
\pagerange{\pageref{firstpage}--\pageref{lastpage}}
\maketitle

\begin{abstract}
Giant pulses (GPs) occur in high magnetic-field millisecond pulsars (MSPs) and young Crab-like pulsars. Motivated by the fast radio bursts (FRBs) discovered in a globular cluster (GC) in the M81, we undertook baseband observations of PSR J1823$-$3021A, the most active GP emitter in a GC with the MeerKAT UHF band receiver (544-1088 MHz). The steep spectral index of the pulsar yields a GP rate of over 37,000 GPs/hr with $S/N>7$, significantly higher than the 3000 GPs/hr rate detected by Abbate et al. 2020 with the L-band (856-1712 MHz) receiver. Similarly to Abbate et al 2020, we find that the GPs are (1) strongly clustered in 2 particular phases of its rotation, (2) well described by a power-law in terms of energies, (3) typically broadband, and have steep spectral indices of $\approx-3$. Although the integrated pulse profile is not significantly polarised ($<1\%$ linear and $<3\%$ circular), one of the brightest GPs displays notable polarisation of 7\% (linear) and 8\% (circular). The high-time resolution data reveals the GPs have a range of single-peak and multi-peak morphologies, including GPs with three distinct peaks. For the first time, we measured the temporal scattering of the pulsar using 49 bright, narrow, single-pulse GPs, obtaining a mean value of $5.5\pm0.6$ $\mu$s at 1 GHz. The distinct periodicity and low polarisation of GPs differentiate them from typical FRBs, although potential quasi-periodicity substructures in some GPs may suggest a connection to magnetars/FRBs.

\end{abstract}




\section{Introduction}
Giant pulses (GPs) are intense radio flashes generated by pulsars that significantly exceed their mean flux levels. There is no strict definition of how much the flux density should be higher than the mean pulse flux to be considered a GP, although a factor of 10 is widely used. 
GPs have been primarily seen arising from young pulsars and millisecond pulsars (MSPs) \citep{Knight2005}, such as PSR B1937+21 \citep{Cognard1996, Soglasnov2004}, PSR B1957+20 \citep{Main2017}, PSR B1821$-$24 \citep{Romani2001}, PSR J0218+4232 \citep{Knight2006a}, PSR J1824$-$2452A \citep{Knight2006b} and PSR J1823$-$3021A \citep{Knight2007}. Among them, PSR J1824$-$2452A and PSR J1823$-$3021A are the only known GP-emitting MSPs that reside in globular clusters (GCs).

PSR J1823$-$3021A resides in the core of the bulge GC NGC 6624 which lies at a distance of 7.9 kpc and has a metallicity of [Fe/H] $=-0.44$ \citep{Harris2010}. So far, 12 pulsars have been found in this GC \citep{Biggs1994, Lynch2012, Ridolfi2021, Abbate2022}. PSR J1823$-$3021A was the first of these to be discovered and is a solitary MSP with a period of 5.44 ms and a dispersion measure (DM) of 86.89 pc cm$^{-3}$ \citep{Biggs1994}. It has a ``characteristic age'' (based on dipole radiation) of 25 Myr and has a moderate magnetic field strength of 4$\times$10$^{9}$ G, which is high for MSPs.

\citet{Knight2005} was the first to detect GPs from PSR J1823$-$3021A, detecting 5 GPs at 685 MHz and 14 at 1405 MHz with the Parkes/Murriyang radio telescope, using a GP definition of $S/N >9.5\sigma$. They detected GPs with pulse energies of up to 64 times the mean pulse energy. In a follow-up study, \citet{Knight2007} conducted a 19,200 s ($\approx5.3$ hr) observation of PSR J1823$-$3021A at a centre frequency of 685 MHz, also using Parkes/Murriyang, and they detected a total of 120 GPs. The pulsar is thus an excellent target for the significantly more sensitive MeerKAT telescope, and \citet{Abbate2020} performed a total of 5 hr (2 observations of 2.5 hrs each) observation of PSR J1823$-$3021A in L-band (856-1712 MHz) in late 2019, detecting 14,350 GPs with a $S/N >7$. The pulses were found to occur at phases similar to the ordinary emission and to have a steep distribution of energies in a power-law with index $-2.63 \pm 0.02$. The spectral index of the GP emission was also found to be a steep power-law with slope $-2.81 \pm 0.02$. Consequently, \citet{Abbate2020} predicted that MeerKAT observations in the UHF band (544-1088 MHz) (using the entire 64 antenna array) for the time duration of 5 hr would detect GPs 3.5 times weaker than their L-band threshold, leading to an $\approx 8.5\times$ increase in the GP detection rate. Furthermore, MeerKAT's UHF band is remarkably clean, with only a few remnant mobile phone-related transmissions \citep{Bailes2020} that only affect $\approx10\%$ of the frequency bins. As a result, radio frequency interference (RFI) removal is easier in the UHF band than in the L-band. 

The CHIME team discovered a repeating FRB (FRB20200120E) coming from a source in a GC in the nearby spiral galaxy M81 \citep{Kirsten2022}. Several models suggest GP-emitting pulsars could be the source of the FRBs \citep[e.g.,][]{Lyutikov2016, Li2024}. Being the most active GP-emitting MSP discovered so far, the study of PSR J1823$-$3021A may lead us to a connection between GPs and FRBs. Furthermore, \citet{Geyer2021} discovered a very bright GP ($S/N > 400$) from the PSR J0540$-$6919 which resides in the Large Magellanic Cloud ($\approx50$ kpc away from us). This GP could be seen at $S/N=10$ up to 0.3 Mpc from us, noting the distance of the M81 FRB is $\approx3.6$ Mpc. This suggests the possibility that the FRB from M81 was a very bright GP in an extragalactic GC. These considerations have motivated our UHF observations of PSR J1823$-$3021A with MeerKAT. Also, MeerKAT offers baseband mode at the UHF which allows us to achieve the best time and frequency resolution.

The paper is structured as follows. We describe the observation and data analysis in Section \ref{sec:obana}. The results are presented in Section \ref{sec:results}. The discussion are in Section \ref{sec:dis}. Lastly, we have our conclusions in Section \ref{sec:con}.

\section{Observations and data analysis} \label{sec:obana}
The Pulsar Timing User-Supplied Equipment (PTUSE) backend \citep{Bailes2020} was used to observe PSR J1823$-$3021A at the MeerKAT radio telescope in South Africa \citep{Jonas2009, Booth2012}, in July 2023. The full MeerKAT array consists of 64 antennas, each with a diameter of 13.5 m, 58 of which were available for the observations reported here. The observations are part of the MeerTIME\footnote{http://www.meertime.org} large science project \citep{Bailes2016}. 
Observations were conducted with a centre frequency of 816 MHz across 544 MHz of bandwidth with all Stokes parameters recorded. Further information on the data acquisition and the system can be found in \citet{Bailes2020}. We observed in baseband mode meaning that full voltage data were recorded, allowing us to perform analysis of the pulsar in the highest possible time and frequency resolution of the instrument. The pulsar was observed in tracking mode for 2984 seconds ($\approx50$ minutes) on 19 July 2023. We adopt a system equivalent flux density (SEFD) of 8.0 Jy when determining the flux densities of the GPs, based on the single dish SEFD at 850 MHz with the number of dishes for the observation (58). 

The data were coherently dedispersed at the known DM of PSR J1823$-$3021A (86.89 pc cm$^{-3}$) and divided into single-pulse archives using \texttt{dspsr} \citep{vanStraten2011} using the topocentric period of the pulsar (5.4400 ms) from the ATNF pulsar catalogue \citep{Manchester2005}. Frequency-scrunched versions of each pulse were made from the single-pulse archives for further analysis. The archives were useful for later checking for e.g. RFI, frequency coverage etc for each detected GP. Although 1.88 $\mu s$ is the native time resolution of the data, \texttt{dspsr} can only create archives with $2^{n}$ bins. Therefore, the best time resolution we can achieve is 2.65 $\mu s$ (2048 bins). We did not create full-time-resolution archives for most of our analyses (GP search, $S/N$ estimation, multipeak search and time of arrival statistics of the GPs) since the best time resolution (2.65 $\mu s$) turned out to be significantly lower than the scattering time at the top of the band (mean scatter time scale $\approx$ 5.5 $\mu s$ at 1 GHz). We could therefore use 5.3 $\mu s$ as the time resolution for most of the analyses as this significantly saved computation time, power, and disk space. All the archives were created with 1024 phase bins across the pulse period with a time resolution of 5.3 $\mu s$, twice the time resolution of \citet{Abbate2020} (as noted by them, higher time resolution can assist in resolving the internal structure of GPs). Additionally, we create archives with 2.65 $\mu s$ time resolution for the brightest 50 GPs for reliable scattering-time measurements (we discuss the scattering time scale measurements in detail in Section \ref{sec:scattering}). 


RFI was identified using \texttt{xprof} (Bailes in prep.) followed by careful visual inspection of a subset of archives in which GPs did not take place. The UHF band at MeerKAT is not as affected by RFI as L-band where $\approx10\%$ and $\approx50\%$ of the frequency bins are affected by RFI in the UHF and L-band, respectively. Regions of strong narrow-band RFI were identified at 890-897 MHz, 935-960 MHz, 1023-1032 MHz, 1040-1042 MHz and 1051-1053 MHz and these were flagged in all the archives with the \texttt{PSRCHIVE} application \texttt{paz} \citep{vanStraten2011}. The high frequency edge of the data (from 1080-1088 MHz) was also found to contain strong intermittent, impulsive RFI and these channels were removed from all archives. A total of 113 out of 1024 frequency channels were flagged.

The frequency-scrunched single-pulse archives (5.44 ms each) were then used to calculate the signal-to-noise ratio ($S/N$) of the total intensity (Stokes I parameter) with \texttt{xprof} (Bailes in prep.) to identify GPs. \texttt{xprof} measures the $S/N$ of at most one single pulse in an archive (5.44 ms of length). It starts with a 1-bin wide boxcar and gradually increases the number of bins (1 bin at a time until it reaches 3\% of the pulse period). This yields the best width of the boxcar for the optimal S/N. It uses the same algorithm as the psrchive program \texttt{pdmp} for S/N computation, determining the baseline's height and rms and computing the normalised area of the profile under the boxcar. We selected single pulses with $S/N > 10$ as candidate GPs. All candidates were inspected and confirmed as GPs based on the position in phase. We have also manually inspected $O(100)$ GPs near our S/N$\sim$10 threshold and were reasonably satisfied that they were all broadband sources. For candidates in the range $7 < S/N < 10$, there is contamination by a small number of false positives. We discuss this in Section \ref{sec:results}. We use the $S/N$, the width of the GPs and the SEFD as a function of frequency of MeerKAT along with the radiometer equation to scale the peak fluxes of the archives to flux densities (Jy).

\section{Results} \label{sec:results}
\begin{figure}
\centering
	\includegraphics[width=\linewidth]{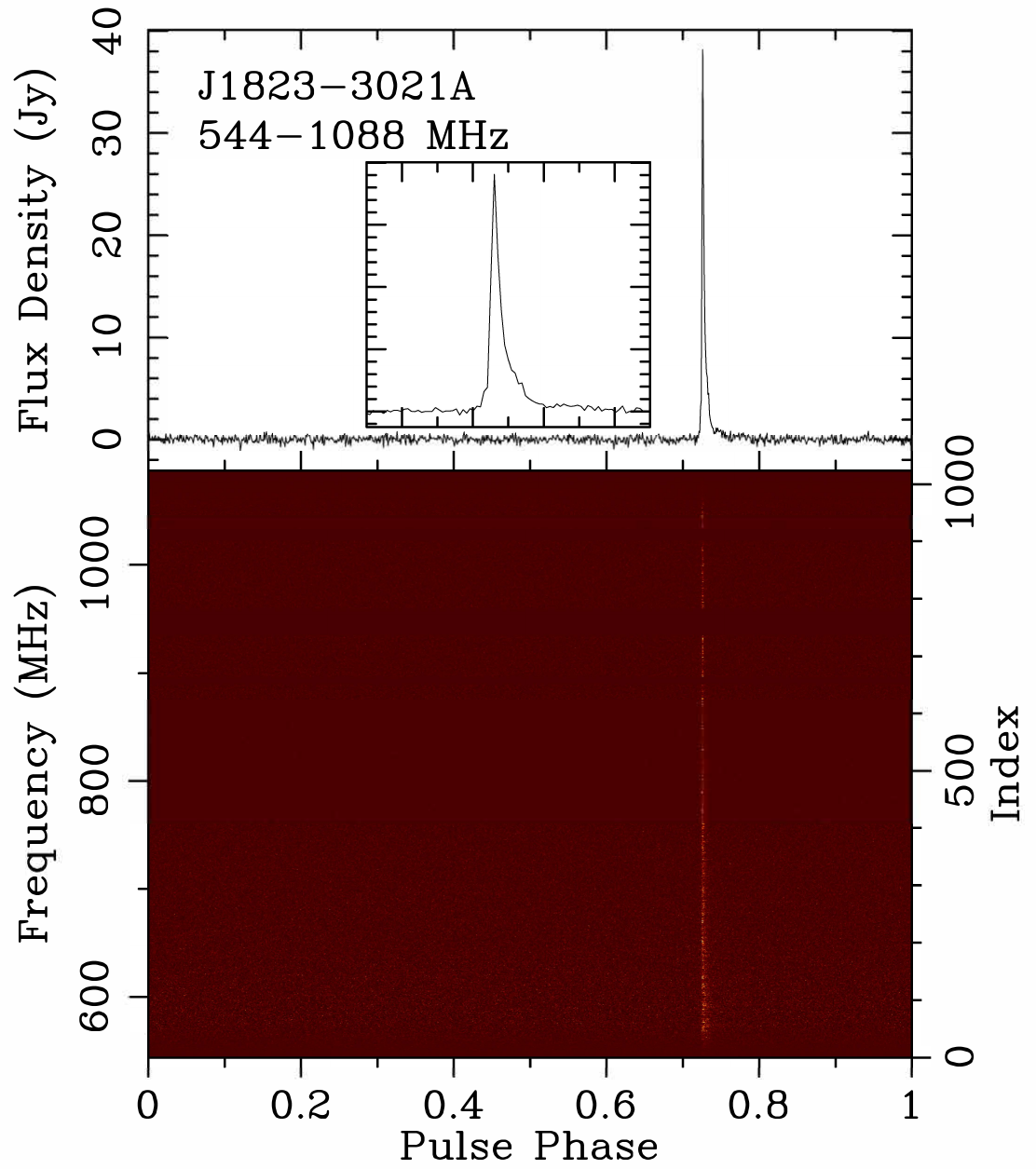}
    \caption{The highest $S/N$ GP detected in the observations. The top panel shows flux density versus phase. The inset shows the zoomed-in of the GP to phase 0.69-0.77. The lower panel shows the waterfall plot of frequency/index versus phase. This GP has $S/N = 254$.}
    \label{brightest.pdf}
\end{figure}

During the 2,984 seconds of observation, 9366 GPs with $S/N > 10$ were detected. The brightest pulse detected has an $S/N \approx 254$ (see Fig. \ref{brightest.pdf}). We estimate a detection rate of $11,300 \pm 100$ GPs/hr for $S/N > 10$. \citet{Abbate2020} obtained a detection rate of about 2870 GPs/hr for $S/N > 7$ and they showed two major clusterings of GPs in the pulsar's rotational phase. In this work, the GPs are also strongly clustered in 2 particular phases of rotation (the C1 and C2 regions), but for $7 < S/N < 10$, our sample GPs are contaminated by a small number of false positives. We found 358 false positives in a phase range of 0.838 (outside the C1 and C2 phase ranges) with $S/N$ between 7 and 10. Using the pulse phase location of the GPs, we estimated the number of false positives to be 358 / $0.838 \times (1-0.838) \approx 69$ in the C1 and C2 phase range between S/N 7-10 for 2984 s. This corresponds to a false positive rate of $69 / (2984 / 3600) \approx$ 83/hr coincident with the range of phase covered by C1 and C2.
To compare with \citet{Abbate2020}, we obtained $31,134$ GPs in 2984 seconds after correcting for the small false positive rate. We estimate a detection rate of $37,000 \pm 200$ GPs/hr.  Thus, the GP rate in the UHF band is 13.5 times higher than the rate in L-band. This is somewhat higher than \citet{Abbate2020}'s prediction of an 8.5 times higher rate.


A comparison of the specifications of the telescopes, sensitivity and results (observing frequencies, SEFDs, time resolution, numbers of detected GPs, and the GP rates) between \citet{Knight2007}, \citet{Abbate2020}, and this work is shown in Table \ref{tab:comparwork}.

\begin{figure*}
\centering
	\includegraphics[width=470pt]{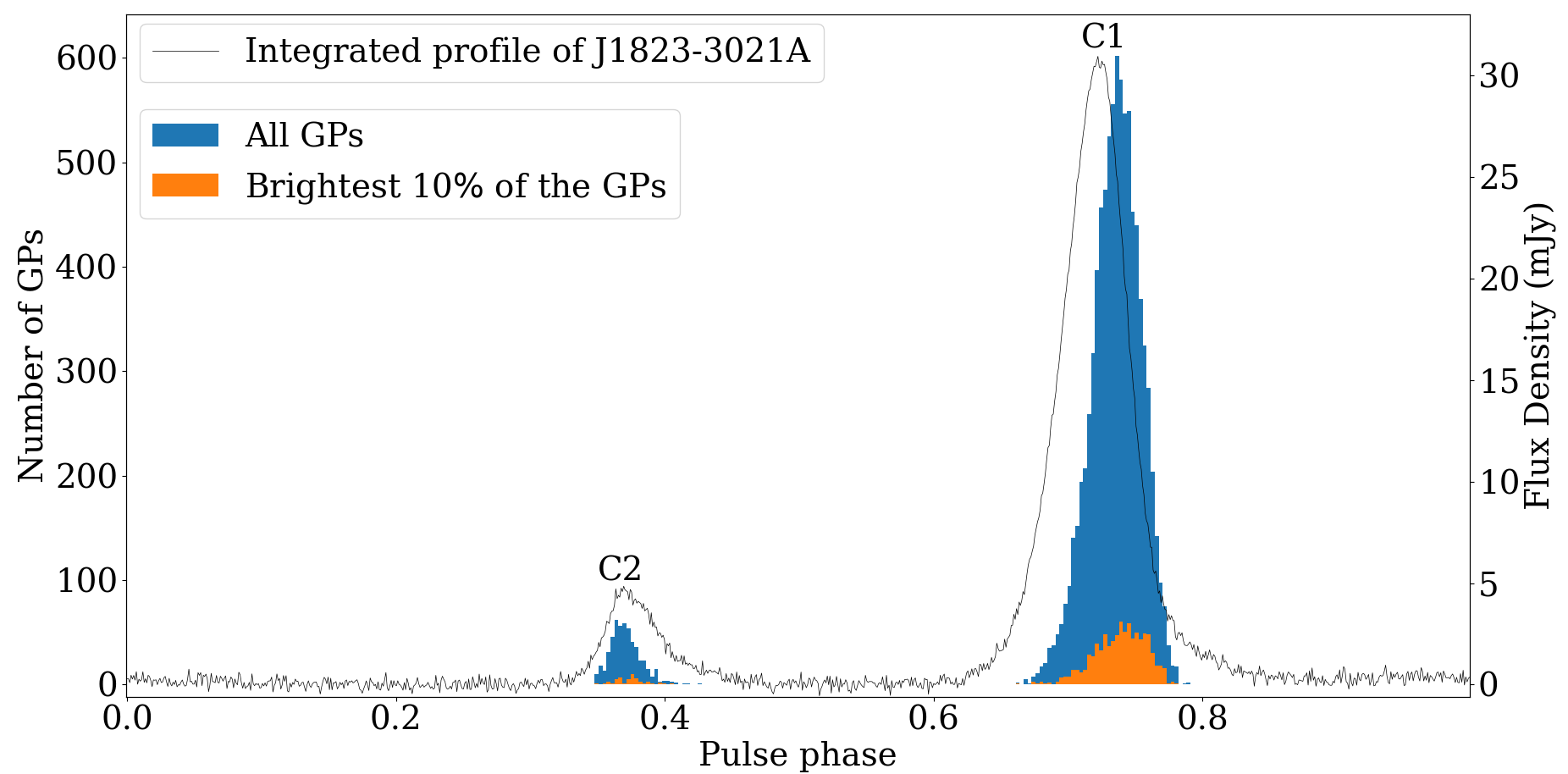}
 
    \caption{The blue histogram shows the phase distribution of GPs with $S/N > 10$. The orange histogram shows the phase distribution of the brightest 10 per cent GPs. The integrated radio emission profile of PSR J1823$-$3021A is shown in black. Note that GPs fall towards the trailing side of component C1.}
    \label{C1_C2.png}
\end{figure*}

We estimate the mean $S/N$ per pulse as 1.6, using the folded profile $S/N$ over the full 2984 s, which gave $S/N =1181$. An S/N of 7 is about 4 times the mean S/N per pulse of the pulsar.  
The GPs mainly fall within the envelope of the folded profile for the total observation, as was seen in both previous studies \citet{Knight2007, Abbate2020}. In Fig. \ref{C1_C2.png}, we show the phase distribution of GPs with $S/N > 10$, superimposed on the integrated radio emission of the pulsar. We have detected the main component C1 and the second component C2 which is the same as \citet{Abbate2020}. However, we do not see any significant detection of the C3 component that was seen in \citet{Abbate2020}. A total of 8837 GPs fall within the main component of the integrated emission C1 with an average of 1 every 0.33s, while only 529 fall within the component C2 with an average of 1 every 5.7s. The window of occurrence has a small dependence on the energy of the GPs as shown in the orange histogram of Fig. \ref{C1_C2.png}. This histogram shows the phase distribution of only the brightest 10 per cent of the GPs and falls towards the trailing side of component C1. 

\begin{table*}
	\centering
	\caption{Summary of the specifications of the observations and the results between \citet{Knight2007}, \citet{Abbate2020} and this work.}
	\label{tab:comparwork}
	\resizebox{\textwidth}{!}{
	\begin{tabular}{llllllll} 
        \hline
         Work & Telescope & Number of antennas & SEFD & Time resolution & Band & Number of GPs detected & Detection rate\\
        \hline
        \citet{Knight2007} & Parkes/Murriyang & Single dish & 66 Jy & 16$\mu s$ & UHF band & 120 in 5.3 hr ($S/N>9.5$) & $\approx23$ GPs/hr\\  
        & & & & & (648-712 MHz) & \\ 
        \hline
        \citet{Abbate2020} & MeerKAT & 1st observation: 36 & $\approx11$ Jy & 10$\mu s$ & L-band & 14,350 in 5 hr ($S/N>7$) & $\approx2870$ GPs/hr\\  
        & & 2nd observation: 42 & $\approx10$ Jy & & (856-1712 MHz) & \\  
        \hline
        This work & MeerKAT & 58 & $\approx8$ Jy & 5.3$\mu s$ & UHF band & 31,134 in 0.8 hr ($S/N>7$) & $\approx 37,000$ GPs/hr\\
        & & & & & (544-1088 MHz) & \\
        \hline
	\end{tabular}}
\end{table*}

\subsection{Energy distribution}
\begin{figure}
\centering
	\includegraphics[width=\linewidth]{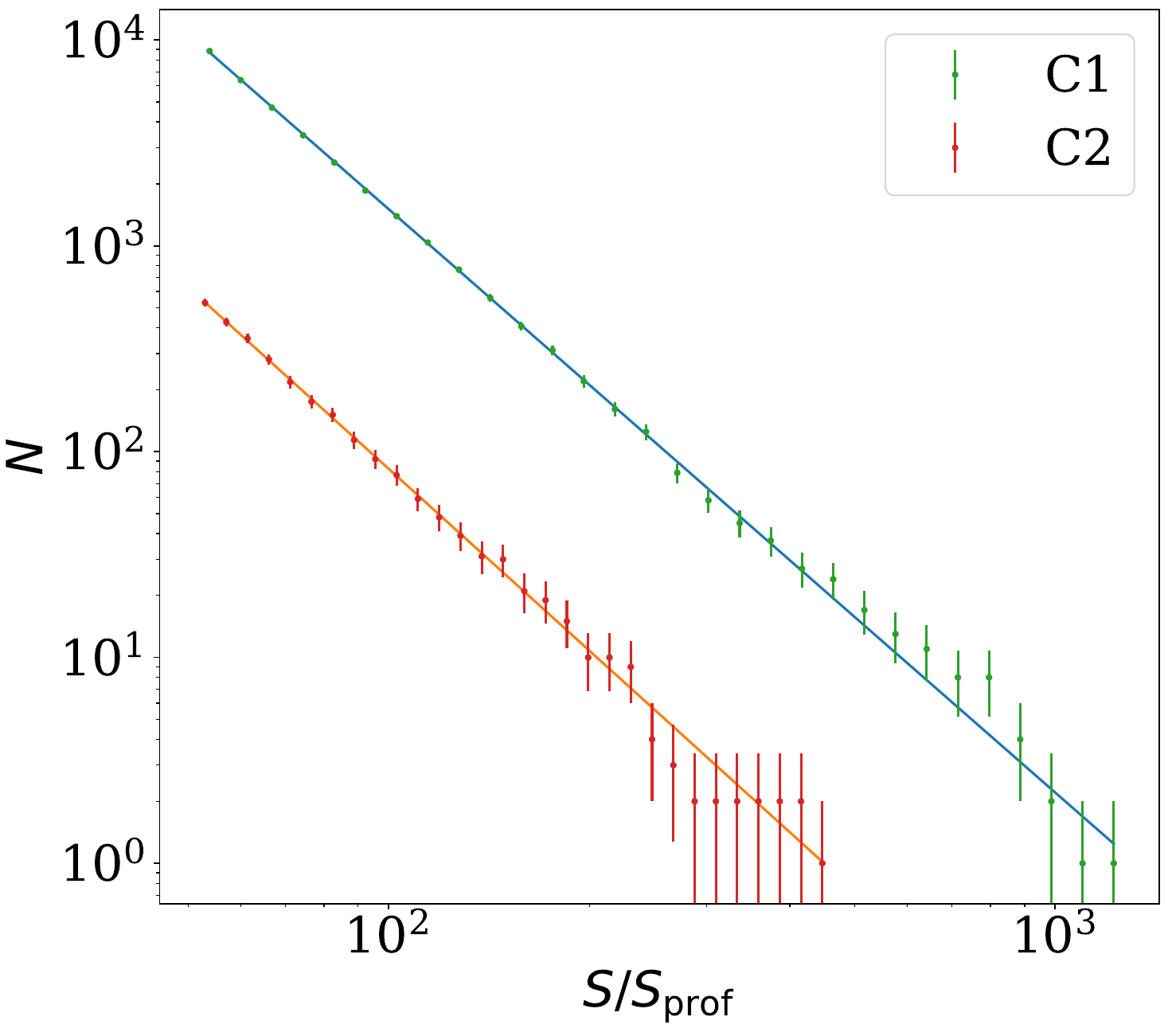}
 
    \caption{Cumulative distribution of the $S/N > 7$ GP energies $S$ in the different components C1 and C2 normalised by the average pulse energy $S_{\text{prof}}$. Note that both axes are on a logarithmic scale. The energies of the GPs follow a power-law distribution where the GPs in C1 are fitted with a slope of $-3.02 \pm 0.01$ and the GPs in C2 are $-2.96 \pm 0.03$.}
    \label{powerlaw_7sigma.pdf}
\end{figure}

We use the \texttt{optimize.curve\_fit} function in \texttt{Scipy} \citep{scipy2020}, which uses a non-linear least squares method to fit for the energy distributions of the GPs. As shown in Fig. \ref{powerlaw_7sigma.pdf}, the GPs with $S/N > 7$ follow a power-law distribution. GPs in components C1 and C2 are treated separately. We show the number of GPs as a function of the GP flux density $S$ relative to the flux density of the average radio pulse $S_{\text{prof}}$. The power law fits of the C1 component and the C2 component are shown by the blue line and orange line, respectively. The GPs in C1 are fitted with a power law with a slope of $-3.02 \pm 0.01$, while the GPs in C2 are $-2.96 \pm 0.03$. This power-law index is less steep than the value estimated by \citet{Knight2007} of $-3.1$. In \citet{Abbate2020} they showed a less steep slope of $-2.63 \pm 0.02$ in C1 and $-2.79 \pm 0.04$ in C2. %
The power-law is an excellent fit to the distributions in both cases over the full range of GP energies.

\subsection{Spectral indices of the GPs}
\begin{figure}
\centering
	\includegraphics[width=\linewidth]{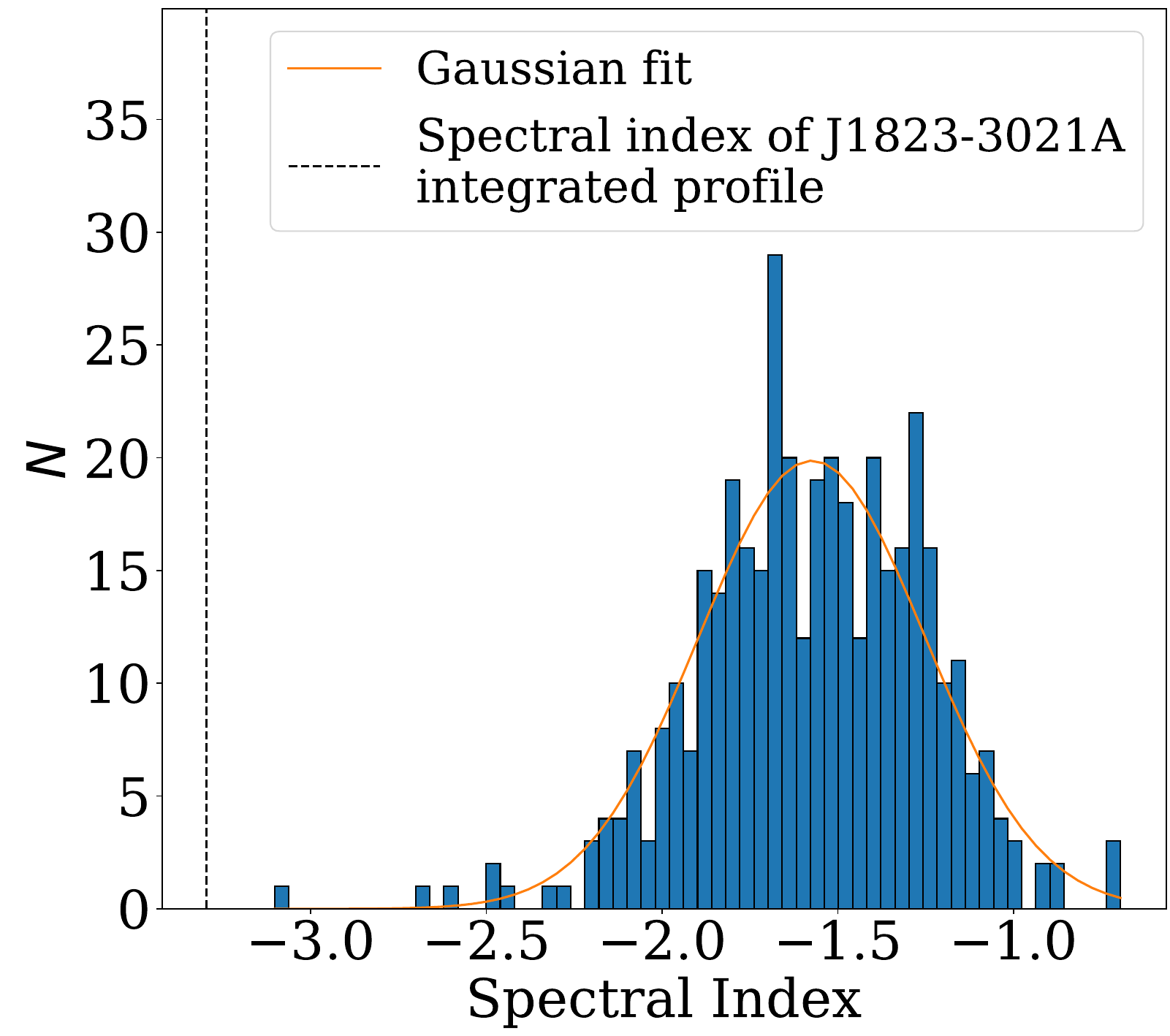}
 
    \caption{Histogram showing the spectral indices of the 400 brightest GPs. The dashed black line shows the spectral index of the integrated profile. The orange line shows the Gaussian fitting of the distribution. The 400 brightest GPs follow a Gaussian and they have less steep spectral indices compared to that of the pulsar integrated profile.}
    \label{spec_hist.pdf}
\end{figure}

To estimate the spectral indices of GPs, we compute their energy as a function of frequency using the radiometer equation, taking into account the sky and system temperatures as a function of frequency (note that the sky temperature varies by a factor of $\approx2$ from 544 MHz to 1088 MHz). Following \citet{Abbate2020}, we fit spectral power-law indices to the spectrum of the GPs. The resulting histogram of the 400 brightest GP spectral indices is shown in Fig. \ref{spec_hist.pdf}. The measured spectral indices follow an approximately Gaussian distribution (orange curve) centred at $-1.6$ with a standard deviation of 0.32 and a median of $-1.57$. We note that the average error on the spectral indices of the GPs is 0.29 which is significantly larger than the measurement errors in \citet{Abbate2020} ($\approx0.09$). This is most likely because the bandwidth in this work (435 MHz) is narrower than that of \citet{Abbate2020} (856 MHz). The spectral index of the integrated profile of the pulsar is $\approx-3.30$ which is shown by the dashed black line. The spectral index of the integrated profile is steeper than the spectral index for the averaged GPs. This was also noted by \citet{Abbate2020} and appears to be a common phenomenon in GP emission \citep{Kinkhabwala2000}.

\subsection{Arrival time statistics}
The exploration of GP arrival times in PSR J1823$-$3021A offers intriguing insights into the underlying statistical distribution. Notably, the emission times of GPs have been found to be well-described by a Poissonian distribution \citep{Lundgren1995, Knight2007}. Within this probabilistic framework, the time intervals between successive GPs follow a consistent and characteristic exponential decay \citep{Abbate2020}. This fundamental property implies that the occurrence of GPs is entirely independent of one another, providing a valuable foundation for the analysis. We create a mean frequency-scrunched pulse profile for the pulsar using the full 2984 s observation. Using this as our standard profile in \texttt{xprof}, we estimate the time of arrival (ToA) of each GP by cross-correlation. The wait time intervals are calculated as the difference of ToAs between consecutive GPs for use in the following analyses.

\begin{figure}
\centering
	\includegraphics[width=\linewidth]{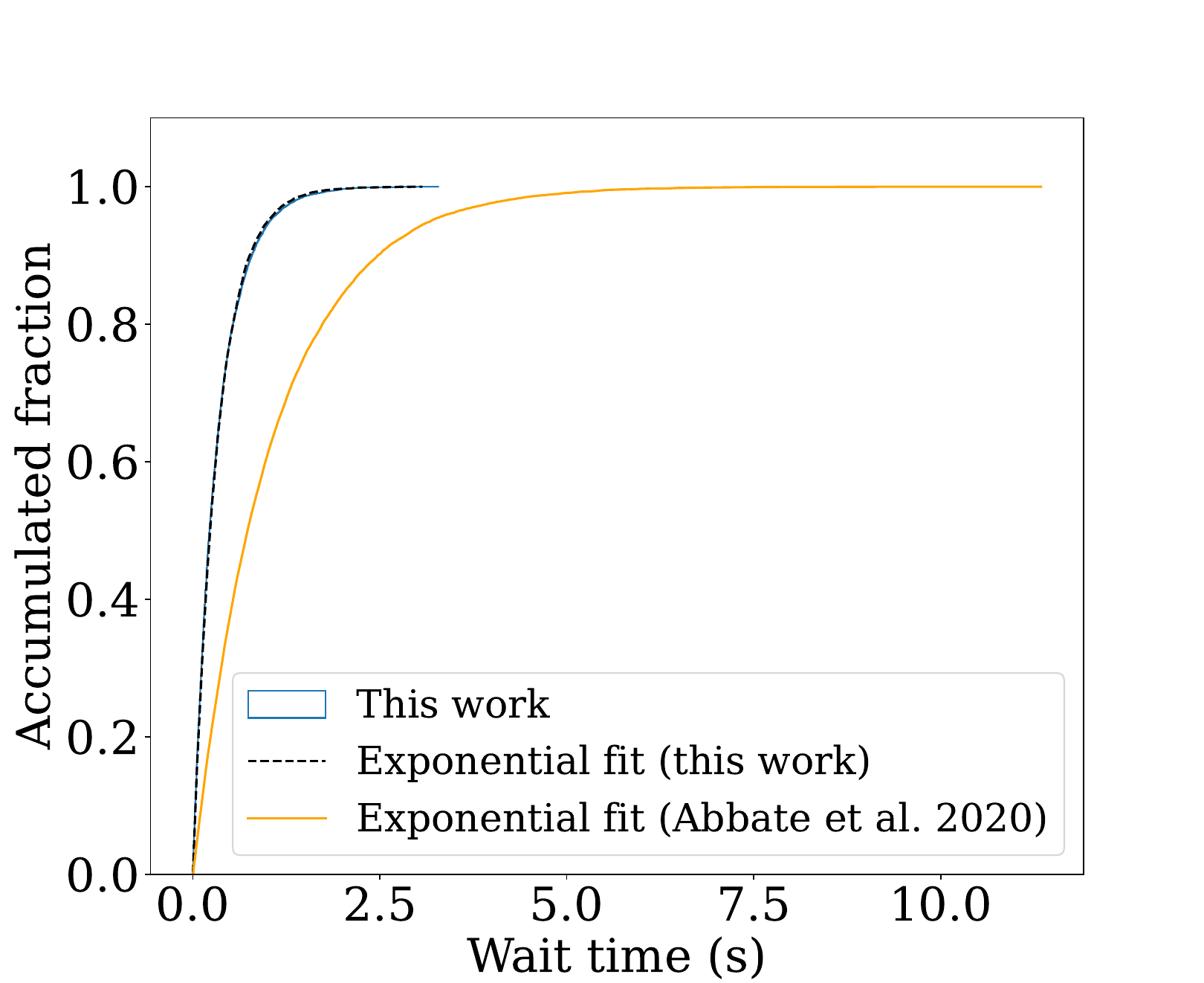}
 
    \caption{Normalised cumulative distribution of the time interval between GPs with $S/N > 10$ is shown by the blue histogram. The black dashed line shows the exponential fit from this work. The yellow curve shows the exponential fit from \citet{Abbate2020}. The exponential curve from this work (decay time of exponential $\approx3.019$s) is steeper than the previous work (decay time of exponential $\approx1.009$s) due to a higher detection rate in the UHF band compared to the earlier study in the L-band.}
    \label{cumul_time.pdf}
\end{figure}

Fig. \ref{cumul_time.pdf} shows the cumulative distribution of wait time intervals between GPs with $S/N > 10$, complemented by integrating the best-fitting exponential distribution where an exponential fit accounts for randomised events.  The cumulative distribution of time intervals between GPs shown by the blue histogram is well-fitted by an exponential (dashed line). The decay time of the fitted exponential is $\approx3.019$s, larger than the one ($\approx1.009$s) from \citet{Abbate2020} (solid orange curve) due to the higher GP rate in the UHF band compared to the L-band. There are also events with longer wait times between two consecutive GPs in \citet{Abbate2020}. 

\begin{figure}
\centering
	\includegraphics[width=\linewidth]{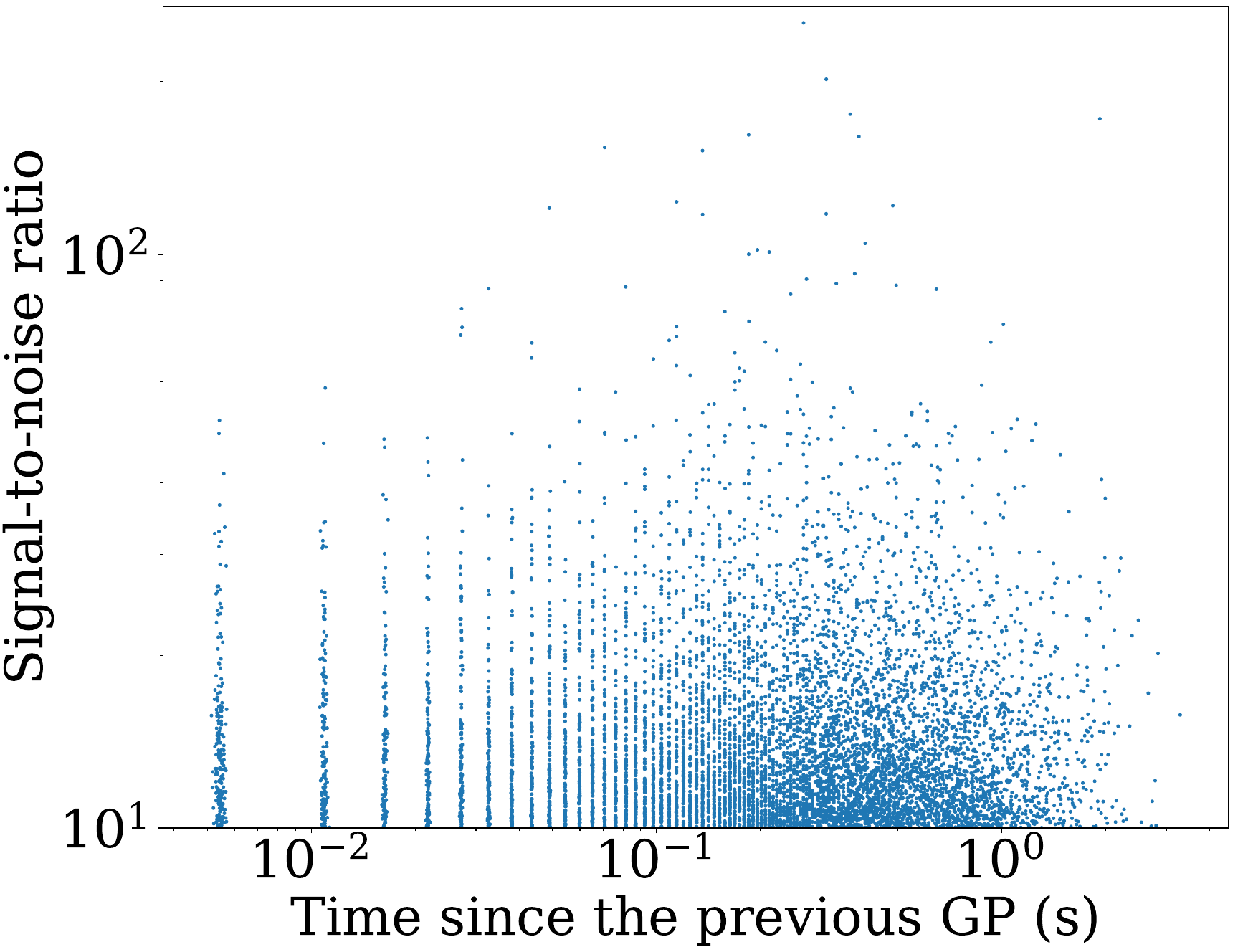}
 
    \caption{Value of the $S/N$ of each GP in the C1 component plotted against the time that has elapsed from the previous GP. The vertical striping is due to integer multiples of the 5.44 ms period of the pulsar. Here we only show GPs with $S/N > 10$.}
    \label{time_snr.pdf}
\end{figure}

In a separate analytical perspective offered in Fig. \ref{time_snr.pdf}, the correlation between the $S/N$ of individual GPs in the C1 component and the wait time since the previous GP is examined. The clusters of data points in vertical stripes show a significant periodicity of the GPs which each strip is 5.44 ms (the period of the pulsar) separated. No discernible correlation emerges from this analysis which shows similar results as \citet{Abbate2020}. This suggests that GP intensity may be inherently independent of the cumulative energy accumulated in the pulsar's magnetosphere after a previous GP \citep{Abbate2020}. 

\subsection{GPs with multiple components} \label{sec:multi}
\begin{figure}
\centering
        \includegraphics[width=\linewidth]{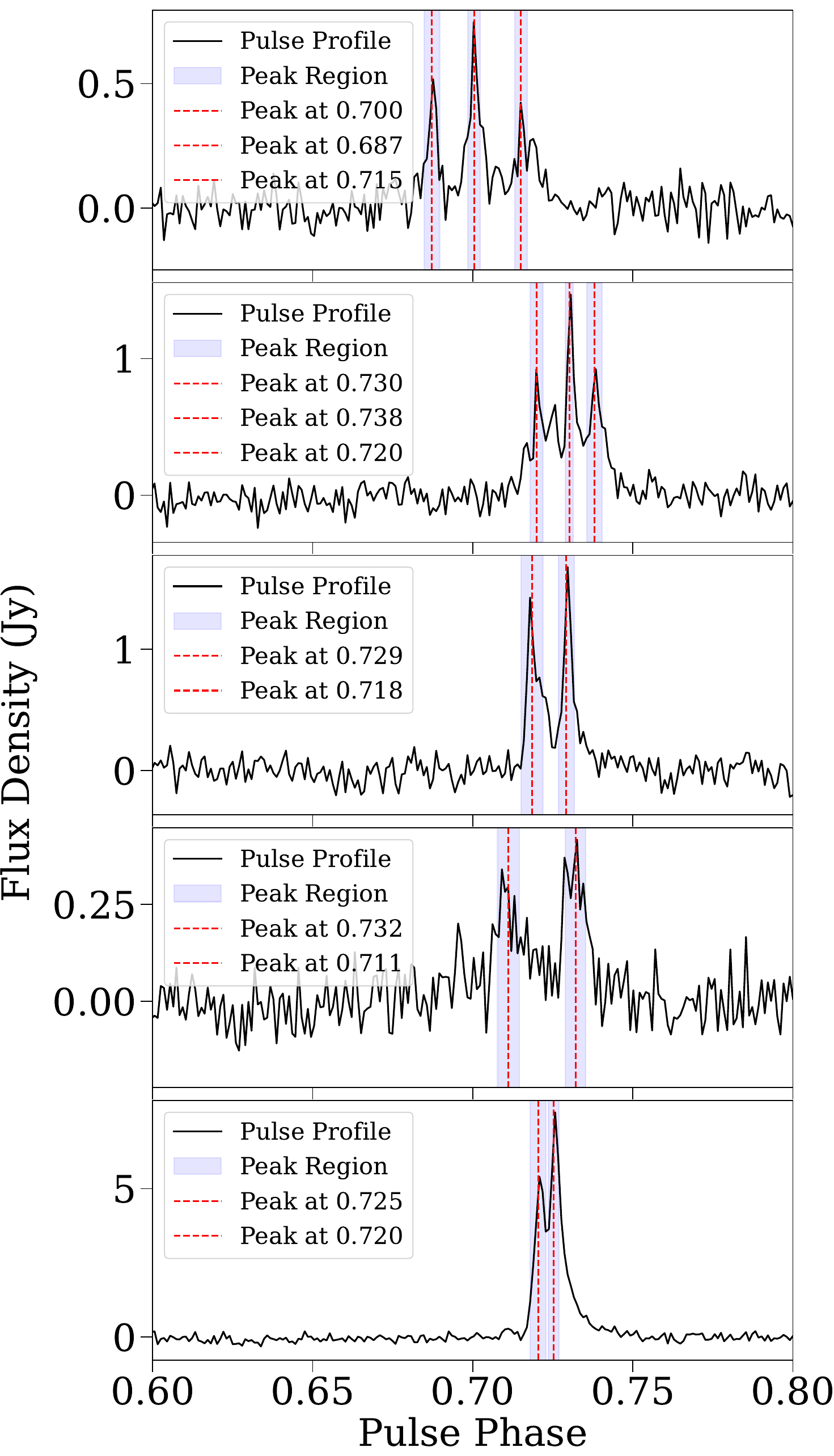}
    \caption{Examples of GPs showing multiple components at the C1 position. The pulse is zoomed in to the pulse phase of 0.60-0.80. The blue shaded regions indicate where $S/N$ is measured for each sub-peak, and the red dashed lines indicate the middle point of the region.}
    \label{multi_1.pdf}
\end{figure}

\begin{figure}
\centering
	\includegraphics[width=\linewidth]{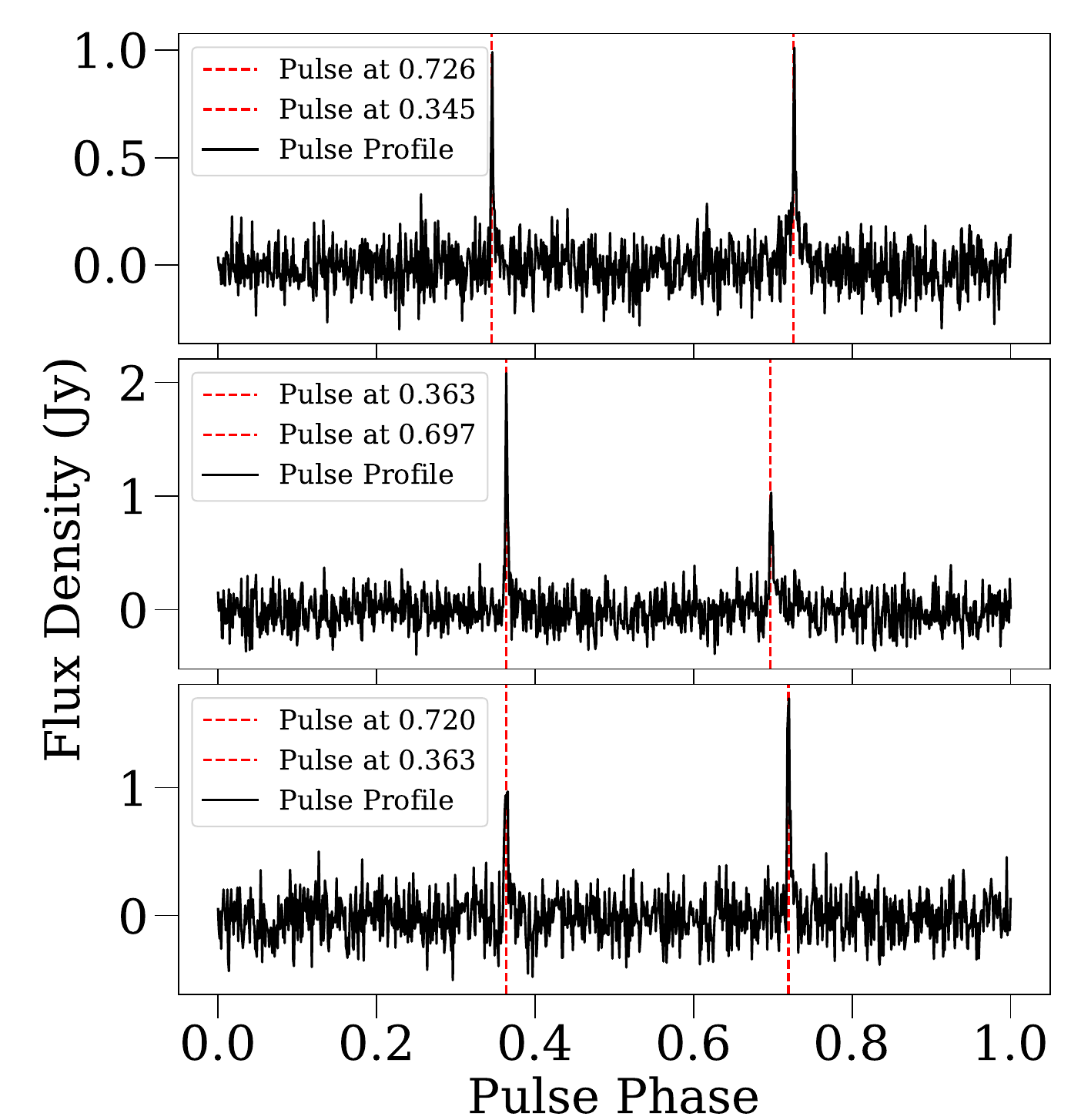}
    \caption{Examples of GPs showing multiple components at both C1 and C2 positions, and occurring within a single rotation of the pulsar. The red dashed lines indicate the positions of the components.}
    \label{multi_2.pdf}
\end{figure}

A significant number of GPs exhibit multiple component profiles. 
These fall into two types. The first type appears as multi-peak structures in one component. We perform detailed single peak S/N measurements to the 9366 GPs mentioned in Section \ref{sec:results}. We first locate up to 5 local maxima within the phase coverage of the archive. Note that each maximum should be separated by at least 10.6 $\mu s$ to the other maxima, which is twice our time resolution (following \citet{Abbate2020}, who also used twice their time resolution).  Then, for each local maximum, search either side of the peak position for the flux to drop to half of the peak flux. Pixels locations 1 pixel less and 1 pixel more than this range are used to measure the S/N of the sub-peak. This also defines the start and end points of the peak. We then measure the $S/N$ within the peak region (shaded blue in Fig. \ref{multi_1.pdf}). We consider a sub-peak to be significant only if its S/N is greater than 10 and its peak is 3-sigma higher than the defined start and end points (to avoid mistaking noise within a GP for a multi-peak component).  The peak centre (marked by the red dashed lines in Fig. \ref{multi_1.pdf}) is the point halfway to the peak region. 
Since our multi-peak GP search algorithm relies on a minimum separation between peaks, it will miss the GPs with multiple components in which peaks overlap (e.g., two components but with only one tip of the peak visible). Merged multiple component GPs are less straightforward to analyse as the scattering tail needs to be accounted for as well (we leave this for future work using coherent descattering techniques following \citet{Main2017}). We focus here on multiple peak GPs with obvious tips and sufficient S/N of each of the peaks. After running the multipeak search algorithm, we found 119 double-peak and 4 triple-peak GPs having $S/N > 10$ where all occur at the C1 position. Multi-peak GPs with three peaks are seen for the first time due to our improved time resolution. Examples of this type of GP are shown in Fig. \ref{multi_1.pdf}. The other type is when GPs of both C1 and C2 occur within the same period of the pulsar: we found 18 of these in which both C1 and C2 components exceeded an $S/N$ of 10. Examples are shown in Fig. \ref{multi_2.pdf} where the pulse phases align with those depicted in Fig. \ref{C1_C2.png}. We thus find that $\approx 1\%$ of the GPs manifest multiple structures of these two types.
Examples of all these events were seen in  \citet{Abbate2020}. New in this work is the presence of GPs with triple peaks, likely due to the higher time resolution of 5.3 $\mu$s (Fig. \ref{multi_2.pdf}). Note that the GPs with multiple components are a sub-sample of the GPs overall. We do not re-count the individual peaks and add them to the number of GPs (i.e. a clump of closely-spaced peaks is still considered to be one GP.)

We estimate the probability of two GPs occurring within a single rotation if we assume the peaks are independent of one another. The probability $P(C1)$ of a single GP occurring at the C1 position in any rotation (C1 in Fig. \ref{C1_C2.png}) $P(C1)$, is $8837/(2984/0.00544)=0.016$ (where we have a total of 2984 s of data). The probability of a single GP occurring at the C2 position in any rotation (C2 in Fig. \ref{C1_C2.png}), $P(C2)$, is $529/(2984/0.00544)=0.00096$. The probability of having two GPs in the same rotation is therefore $P(C1)P(C2) = 1.5 \times 10^{-5}$. With an observation time of 2984 s (which corresponds to 548,529 rotations of the pulsar). We therefore expect about 8 events of GPs happening at both C1 and C2 within a single rotation. Observationally, we find 18 such events taking place within a single rotation. Hence, the presence of C1 and C2 pulses may weakly correlate (at about 3-$\sigma$ level of significance). 

We can use binomial statistics to estimate the number of expected double-peak and triple-peak GPs, with the equation\[
C(N_{\mathrm{C1}}, N_{\mathrm{peak}}) \cdot \left( \frac{1}{N_{\mathrm{rot}}} \right)^{N_{\mathrm{peak}}-1} \cdot \left( 1 - \frac{1}{N_{\mathrm{rot}}} \right)^{N_{\mathrm{C1}} - N_{\mathrm{peak}}},\] 
where $C$ is the binomial coefficient function, $N_{\mathrm{C1}}$ is the number of GPs in the C1 position, $N_{\mathrm{peak}}$ is the number of expected peaks and $N_{\mathrm{rot}}$ is the number of rotations of the pulsar in this observation. Using $N_{\mathrm{C1}} = 8837 $, $N_{\mathrm{rot}} = 548529$ and $N_{\mathrm{peak}} = 2$, we expect 70 double-peak GPs in our sample with $S/N$ > 10. We expect only 0.4 triple-peak GPs ($N_{\mathrm{peak}} = 3$) with $S/N$ > 10. Observationally, we find 119 double-peak GPs and 4 triple-peak GPs. The observed number of multi-peak GPs is significantly more than expected from chance coincidence alone. Therefore, we conclude these multi-peak GPs are more likely to be nanoshot-like substructures \citep[e.g.,][]{Jessner2010, Lin&vanKerkwijk2023} happening in a GP. Interestingly, these substructures are also commonly found in most FRBs \citep[e.g.,][]{Pastor-Marazuela2023}.


\subsection{Polarisation} \label{sec:polar}
\begin{figure}
\centering
	\includegraphics[width=\linewidth]{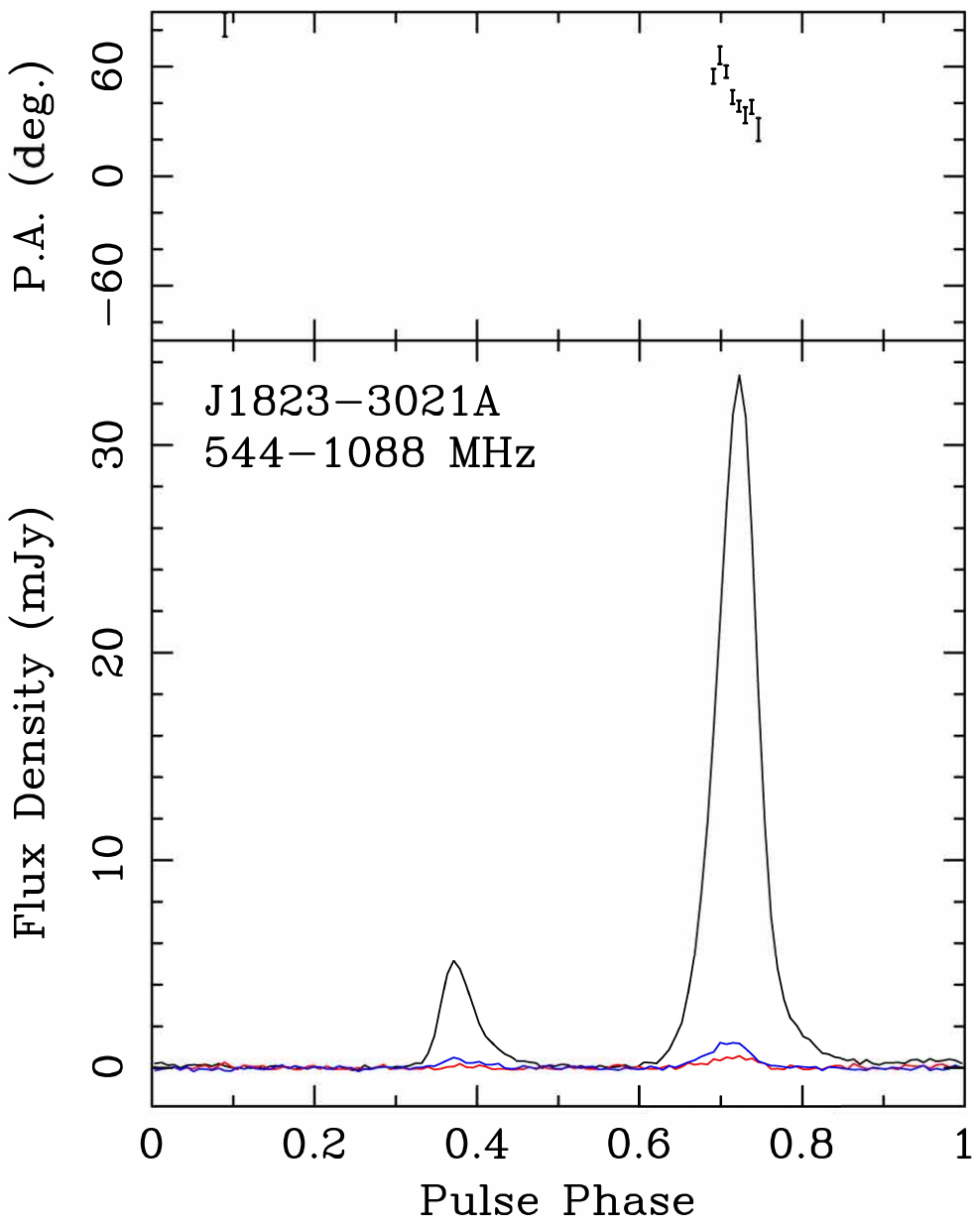}
 
    \caption{The polarisation profile of PSR J1823$-$3021A. In the lower panel, the black line shows the total intensity, the red line shows the linear polarisation and the blue line shows the circular polarisation. Components C1 and C2 show linear polarisation and circular polarisation only at about 1\% level and about 3\% level, respectively. The position angle (PA) measured for the bins is shown in the upper panel.}
    \label{intpro_pol.pdf}
\end{figure}

Pulsar polarimetry with MeerKAT is discussed in \citet{Bailes2020} and \citet{Serylak2021}. \citet{Abbate2020} reported that the emission from PSR J1823$-$3021A is not strongly polarised. 
Full polarisation data were recorded for this observation
and we performed a similar analysis as \citet{Abbate2020} on the polarisation profile of the pulsar. The integrated polarisation profile is shown in Fig. \ref{intpro_pol.pdf} where weak linear (red) and circular (blue) polarisation are seen. Components C1 and C2 show linear polarisation only at about 1\% level and circular polarisation at about 3\% level which agrees with \citet{Abbate2020}. 
We used \texttt{rmfit} \citep{vanStraten2011} obtaining an RM of the $-19.3 \pm 4.1$ rad m$^{2}$, which is consistent within errors of $-16 \pm 3$ rad m$^{2}$ obtained by \citet{Abbate2020}.

\begin{figure}
\centering
	\includegraphics[width=253pt]{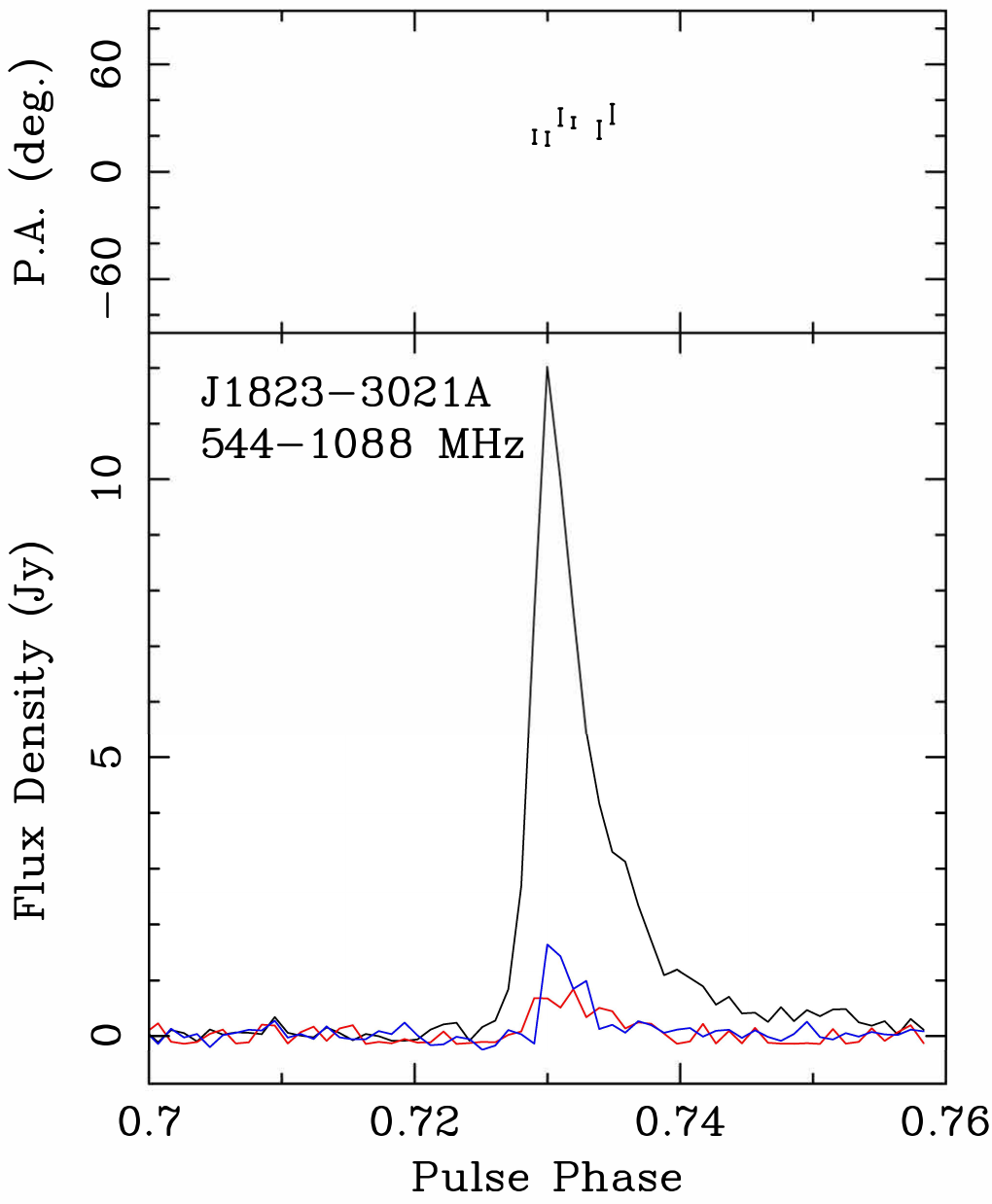}
 
    \caption{The polarisation profile of the most polarised GP with $S/N$ of 203. In the lower panel, the black line shows the total intensity, the red line shows the linear polarisation and the blue line shows the circular polarisation. It shows linear polarisation and circular polarisation at about 7\% level and about 8\% level of the total emission, respectively. The PA measured for the bins is shown in the upper panel. The pulse is zoomed to the pulse phase of 0.70-0.76.}
    \label{pol_GP.pdf}
\end{figure}

\citet{Abbate2020} analysed the polarisation profile of GPs with multiple components and they found that only a few of these GPs are significantly polarised. In this work, we check the polarisation profile of the GPs described in Section \ref{sec:multi} but we do not see significantly polarised GPs with multiple components. Instead, we found most of the more significantly polarised GPs are the brightest ones with high $S/N$. We show the most polarised GP in Fig. \ref{pol_GP.pdf}. It shows linear polarisation at about 7\% level and circular polarisation at about 8\% level of the total emission. It is a very bright GP with an $S/N$ of $\approx203$. We perform \texttt{rmfit} \citep{vanStraten2011} on this GP and estimate an RM of $-24.8 \pm 2.9$ rad m$^{2}$. The RM of this GP is consistent within error with the RM of the integrated profile. We also see that even if the polarisation fraction is quite different (5\% and 6\% different in the circular and linear polarisations), they have similar RMs.

\subsection{Scattering} \label{sec:scattering}
\begin{figure*}
\centering
	\includegraphics[width=455pt]{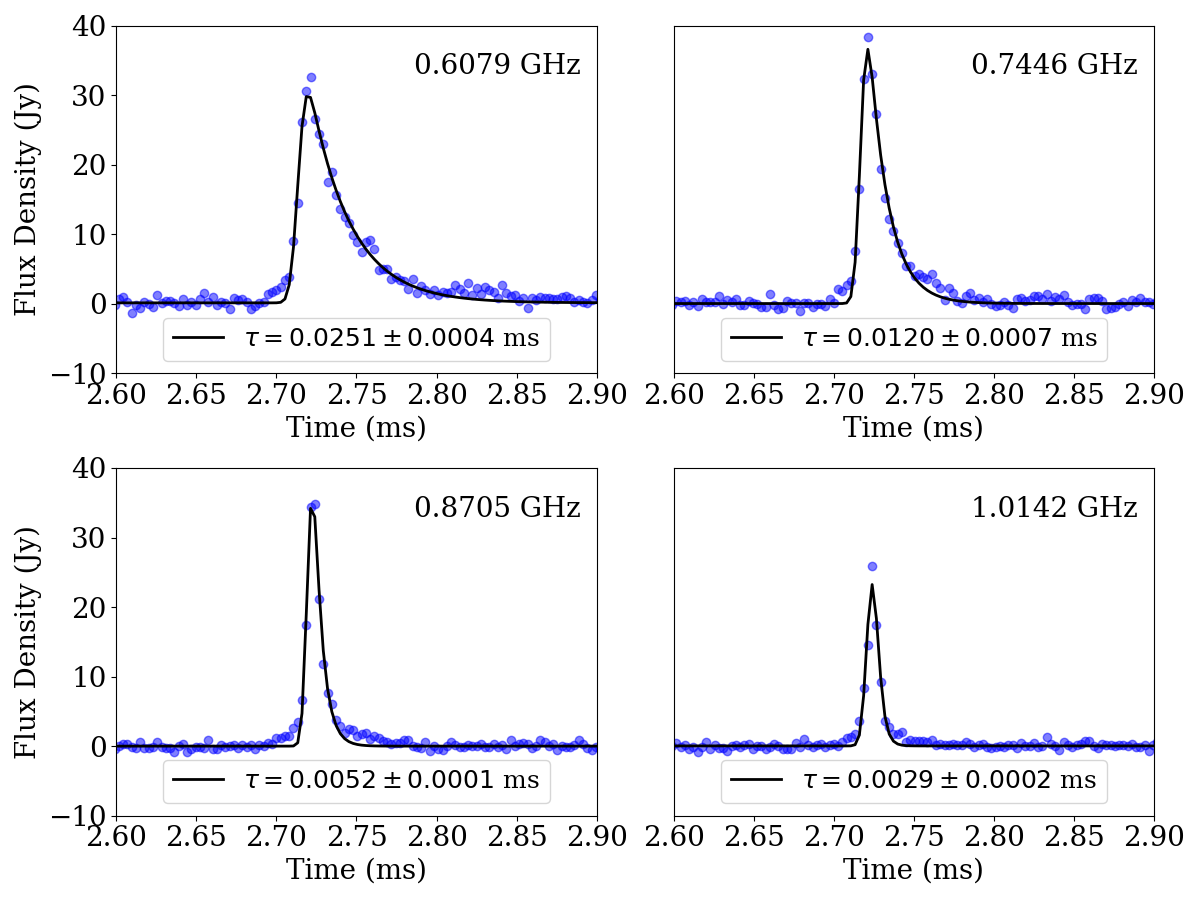}
 
    \caption{An example of the scattering fitting for the brightest GP we have detected ($S/N \approx 254$) at 607.9 MHz, 744.6 MHz, 870.5 MHz, and 1014.2 MHz. The blue dots are the data points of the pulse profiles and the black lines are the fitted models. The results for $\tau$ and $\sigma$ versus frequency for this GP are shown in Fig. \ref{alpha.pdf}.}
    \label{scattering.png}
\end{figure*}

\begin{figure}
\centering
	\includegraphics[width=220pt]{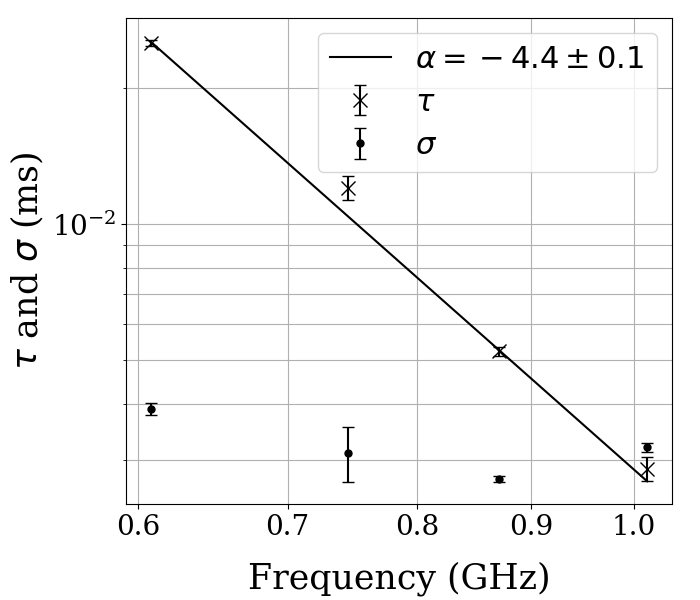}
 
    \caption{The scattering time scale, $\tau$ versus frequency for the brightest GP by fitting in four frequency regions centred at 607.9 MHz, 744.6 MHz, 870.5 MHz, and 1014.2 MHz using the code from \citet{Oswald2021}. Results for $\sigma$ versus frequency is also plotted for reference. The x-axis and y-axis are both in logarithmic scale. The fitted scattering index, $\alpha$ is $-4.4 \pm 0.1$.}
    \label{alpha.pdf}
\end{figure}

Scattering tails have been commonly found in the B0531+21 (Crab pulsar) GPs where the pulses are scattered by a thin screen between the source and the observer \citep{Karuppusamy2010}. As is well known, the exponential scattering tail scales strongly with frequency ($\propto \nu^{4}$) so the effect of scattering is more significant in the lower UHF band. Due to this reason and a better time resolution we have compared to the previous work, we measure the scattering of GPs for this pulsar for the first time. We perform scattering fitting using an open source Python package, \texttt{SCAMP-I} \citep{Oswald2021}. A Gaussian model with a scattering transfer function developed by the algorithm from \citet{Geyer2016} is fitted and later converged using Markov-chain Monte Carlo (MCMC) methods via the software package \texttt{emcee}. We set the fitting method as ISO where it models the screen as an isotropic scatterer. The width of the phase window around a GP for the fitting is set as the period of the pulsar (5.44 ms). We ran the code on the 50 brightest GPs in our sample. The code estimates the scattering timescale ($\tau$) and scattering index ($\alpha$) across four frequency subbands. For 49 GPs the algorithm converged and gave satisfactory parameter and error estimates. The single failed case is a broad GP of width $\approx 32~\mu$s (at 1 GHz), much broader than the scattering timescale, and this GP was removed from the sample for the purposes of the following analysis. The code gives an estimate of the scattering time at 1 GHz. The mean scattering time of the 49 GPs is found to be $\tau = 5.5 \pm 0.6$ $\mu s$. The scattering indices were found to lie in the range $-6.5 < \alpha < -1.1$. This value for the mean $\tau$ is consistent with the scattering arising from the ISM: the NE2001 model \citep{Cordes2002} gives $\tau = 12$ $\mu$s and the YMW model \citep{YMW2017} gives $\tau = 3$ $\mu$s for the DM and position of NGC6624 on the sky. In Fig. \ref{scattering.png}, we show an example of the scattering fitting for the brightest GP we have detected ($S/N$ $\approx$ 254) centred at 607.9 MHz, 744.6 MHz, 870.5 MHz, and 1014.2 MHz. We show the example fit of the scattering index of the brightest GP in Fig. \ref{alpha.pdf} where the fitted $\alpha$ is $-4.4 \pm 0.1$. The width of the pulse, $\sigma$ versus given frequency is also plotted for reference. We show the MCMC corner plots of some example fits in the Appendix \ref{appendix} for visualisation. For the 49 well-fitted GPs, the mean scattering index, $\alpha$ is estimated to be $-2.5 \pm 0.3$  which is higher than the theoretical models predicting $\alpha=$ $-4.0$ (thin screen) or $-4.4$ (Komologorov spectrum). \citet{Geyer2017} also found that pulsars' $\alpha$ are not consistent with the theoretical value of $-4.0$ or $-4.4$.

\section{Discussion} \label{sec:dis}
\subsection{The comparisons with other GP emitters}
We compare the energy distribution, polarisation properties and arrival statistics with PSR J1824$-$2452A, PSR B1937$+$21 and the Crab pulsar. Although PSR J1824$-$2452A is not well-studied, it is the only other MSP GP emitter in a GC. \citet{Knight2006b} found 27 GPs with Parkes at 1.3-3.5 GHz. They fitted the cumulative probability distribution of the GPs. By excluding 6 GPs with the lowest and highest energies, they fitted a power law with a slope of $-1.6$. The P1 (interpulse) and P2 (main pulse) are 72\% and 96\% linearly polarised, respectively. They exhibit no circular polarisation. Their GP sample size is too small to constrain the arrival time statistics. PSR B1937$+$21 is the first MSP to be discovered \citep{Backer1982} and a well-studied active GP emitter. \citet{McKee2019} have studied 4265 GPs from it with the Large European Array for Pulsars. They fitted the pulse energy distribution of the GPs. For the interpulse, they are well-described by a power law with a slope of $-3.99 \pm 0.04$. The main pulses are best described by a broken power law, with slopes of $-3.48 \pm 0.04$ at lower energy and $-2.10 \pm 0.09$ at higher energy. \citet{Mahajan2024} found 60,270 GPs (30,913 GPs/hr) with the Arecibo telescope data at 327 MHz. They fitted the reverse cumulative fluence distribution for the main pulse and the interpulse with power laws separately. The main pulse and the interpulse have a slope of $-1.6$ and $-3.6$, respectively. The main pulse is 81\% and 30\% linearly and circularly polarised, respectively. Generally speaking, the GPs are randomly polarised. Their GPs show an exponential distribution for the arrival time with respect to the main pulse.

The Crab pulsar is a well-studied pulsar for which different power-law indices are reported. \citet{Popov2007} find $-1.7$ to $-3.2$ for main pulse GPs and $-1.6$ for interpulse GPs at 1197 MHz, \citet{Majid2011} find $-1.9$ for the combined main pulse and interpulse distribution at 1664 MHz and \citet{Lin&vanKerkwijk2023} find $-2.0$ for the main pulse and $-1.6$ for the interpulse at 1658.49 MHz. The difference is most likely to arise from the broken power-law-like distribution at high and low energies. \citet{Lin2023} studied 61,998 GPs found by coherently combined European VLBI network observations at 18 cm and they found most GPs are linearly polarised but have little circular polarisation. \citet{Karuppusamy2010} studied the Crab pulsar with the Westerbork Synthesis Radio Telescope (WSRT) wide band at 1311-1450 MHz. With a detection rate of $\approx$13,000 GPs in $\approx$6 hrs, they also found an exponential decay in the distribution of separation times of the GPs. In terms of polarisation, PSR J1824$-$2452A, PSR B1937$+$21 and the Crab pulsar are very different from PSR J1823$-$3021A as they are mostly linear polarised ($>$ 70\%) and exhibit little circular polarisation. PSR J1823$-$3021A is mostly unpolarised where the pulsar profile is only at about 1\% level and about 3\% level in linear polarisation and circular polarisation, respectively. Even for the most polarised GP, it is only at 7\% linear and 8\% circular polarisation. 
To summarise, most other GP emitters have quite different values for their spectral indices that were fitted to the main pulse and the interpulse (a difference of $>$ 0.4). This is quite different to PSR J1823$-$3021A where we have only a difference of 0.06 between the spectral indices of the main pulse, C1 ($-3.02 \pm 0.01$) and the interpulse, C2 ($-2.96 \pm 0.03$). Moreover, broken-power law fits are commonly seen when fitting the spectral index of the main pulse in other GP emitters. However, from both \citet{Abbate2020} and this work, the energy distribution of PSR J1823$-$3021A is well-fitted with a single power law.
The difference in polarisation properties suggests that the GP emission may not be directly related to it. PSR J1823$-$3021A is mostly unpolarised but still very active like PSR B1937$+$21 and the Crab pulsar. The GP emission mechanism may be more related to the magnetosphere of the pulsar instead. In terms of arrival time statistics, PSR J1823$-$3021A shows the same exponential decay in the waiting time distribution as PSR B1937$+$21 and the Crab pulsar. This may suggest that GPs even from different sources are mutually exclusive events which occur randomly.

\subsection{The connection between GPs and FRBs from GCs}
\begin{figure}
\centering
	\includegraphics[width=\linewidth]{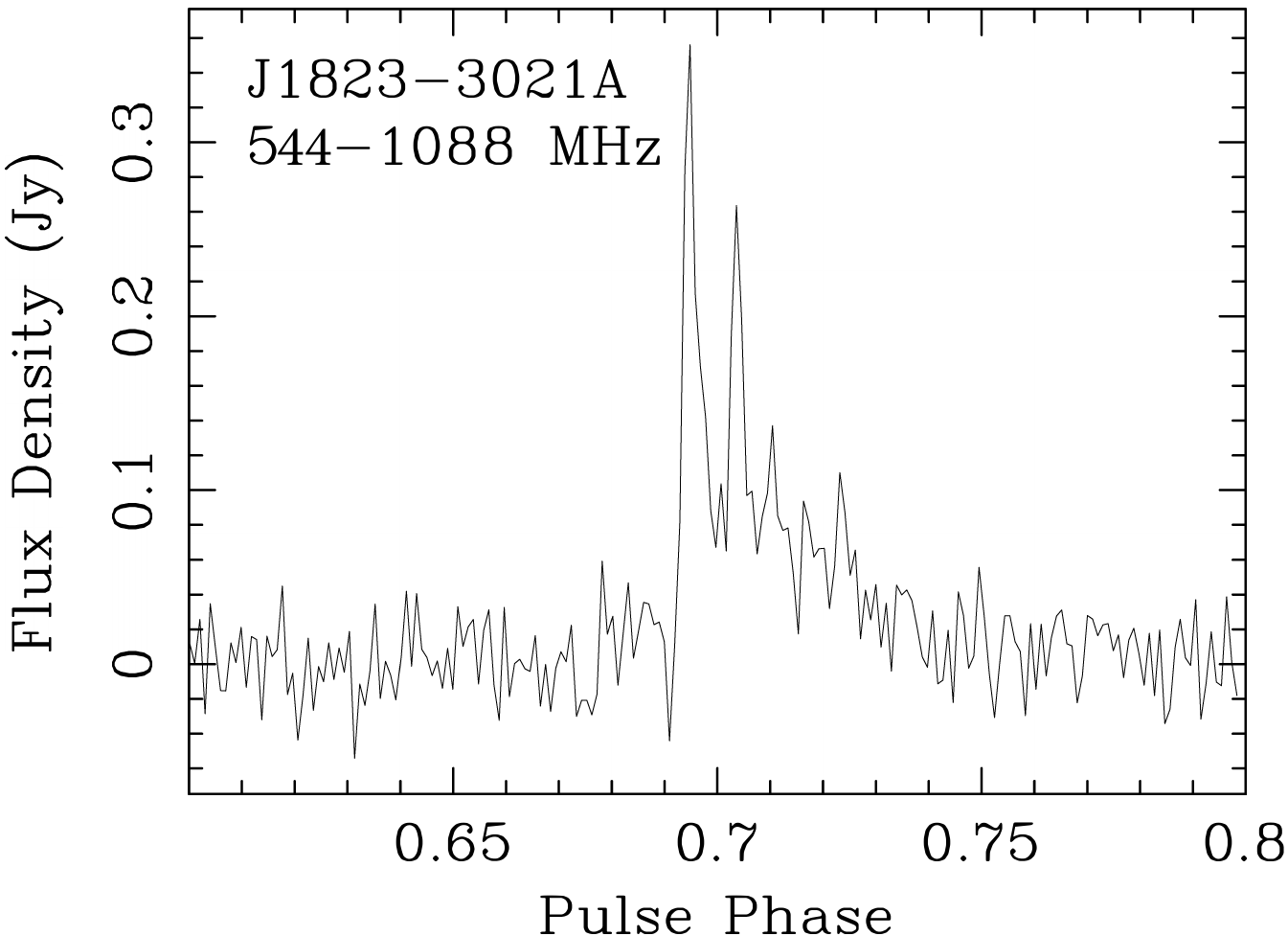}
 
    \caption{An example of a GP that may exhibit quasi-periodicity in the multi-peak components, similar to FRB 20201028 \citep{Pastor-Marazuela2023}. The $S/N$ of this event is $\approx 30$. }
    \label{quasi.pdf}
\end{figure}

\citet{Kirsten2022} discovered a very interesting repeating FRB (FRB20200120E) coming from a source in a GC in the nearby spiral galaxy M81. Several models are suggesting GP-emitting pulsars as the source of the M81 FRB \citep[e.g.,][]{Lyutikov2016, Li2024}. PSR J1823$-$3021A is the most active GP-emitting MSP we have discovered so far. Studying the properties of GC MSPs that emit GPs may also give us clues to a connection to the GC FRBs. In this work, the brightest GP that we have detected is with a $S/N \approx 254$ (Fig. \ref{brightest.pdf}). This GP would still be detectable with MeerKAT with an $S/N$ of 10 if it is $\approx$ 40 kpc away from us. To reach the distance of the M81 FRB ($\approx$ 3.6 Mpc), we need to detect a GP that is about $10^{4}$ times more luminous. \citet{Abbate2020} found a GP with $S/N$ $\approx 370$ in a total of 5 hr observation. We may have found GPs with at least $S/N$ of 370 in the UHF band with a similar length of observation. From Fig. \ref{time_snr.pdf}, we see that there is a strong periodicity in the arrival time of GPs. \citet{Nimmo2023} reported that there is no significant periodicity seen from the burst storm of FRB20200120E. Also, we have not seen any strong periodicity from FRBs so far. However, it is still hard to say if a significant periodicity can be seen from our work if we only considered the brightest GPs (that might be possible to be detected from the extra-galactic universe). Besides, there are sub-millisecond quasi-periodic structures discovered by \citet{Pastor-Marazuela2023} that look similar to what we have found in the multi-peak components in Fig. \ref{multi_1.pdf} and the one in Fig. \ref{quasi.pdf}. \citet{Kramer2024} have found an empirical relation between the rotational period and the time scale of quasi-periodic substructure within single pulses that can be found in every type of radio-emitting neutron star including magnetars which are a favoured model of FRBs. This means quasi-periodicity in GP multi-components may suggest a link to magnetars or FRBs. Whether these multi-peak components seen from the GPs are quasi-periodic is still under investigation and will be discussed in future work. 

Most FRBs are known to be highly polarised \citep[e.g.,][]{Petroff2015, Nimmo2021}. As for FRB20200120E, it is reported that it is 100\% linear polarised \citep{Bhardwaj2021}. In this work, as discussed in Section \ref{sec:polar}, the integrated emission of the pulsar is only slightly polarised (3\% in circular polarisation and 1\% in linear polarisation, Fig. \ref{intpro_pol.pdf}). Even for the most polarised GP (Fig. \ref{pol_GP.pdf}) we have detected, that it is only <10\% polarised in the linear and circular. This suggests a difference between GPs' polarisation profiles and that of FRBs. The baseband nature in our data will enable us to both coherently dedisperse and coherently descatter the large sample of GPs that we found in this work. We intend to publish related work in another future paper.

\section{Conclusions} \label{sec:con}
We have detected 9366 GPs with $S/N > 10$ in 2984s in the MeerKAT UHF band. We estimate a GP rate of $11,300 \pm 100$ GPs/hr with $S/N > 10$ at this frequency. For $7 < S/N < 10$, our sample GPs are contaminated by a small number of false positives. Using the pulse phase statistics, we have estimated the rate of false positives between $7 < S/N < 10$ of $\approx$ 100/hr coincident with the range of phase covered by C1 and C2. To compare with \citet{Abbate2020}, we obtained $37,000 \pm 200$ GPs/hr after correcting for the low false positive rate. Thus, the GP rate in the UHF band is 13.5 times higher than the rate in the L-band ($\approx3000$ GPs/hr). This is somewhat higher than \citet{Abbate2020}'s prediction of an 8.5 times higher rate. This can also be seen from the steeper normalised cumulative distribution of the time interval between GPs for this work compared to \citet{Abbate2020} in Fig. \ref{cumul_time.pdf}. We estimate the mean $S/N$ per pulse as 1.6, using the folded profile $S/N$ over the full 2984s. In Fig. \ref{powerlaw_7sigma.pdf}, the slope of the power law is less steep in our work compared to \citet{Abbate2020}. We find that about 1\% of the GPs reside in the C2 component and the rest in the main component C1 similar to the findings of \citet{Abbate2020}. 
We also studied the arrival time statistics of the GPs. We found similar results as previous studies that the emission times of GPs are found to be well-described by a Poissonian distribution which may mean that GPs are independent of each other. By plotting the integrated polarisation profile of the pulsar, we have seen slight linear and circular polarisation. Components C1 and C2 show linear polarisation only at about 1\% level and circular polarisation at about 3\% level which agrees with \citet{Abbate2020}. Thanks to the high-time resolution of the data, we can detect GPs with multiple components and scattering for the first time. We found that $\approx1\%$ of the GPs have multiple components and a large fraction of GPs with significant scattering tails. The scattering tails of 49 well-fitted GPs have a mean value of $5.5 \pm 0.6$ $\mu$s scaled to 1 GHz and a mean scattering index, $\alpha$ has a value $-2.5 \pm 0.3$ which is higher than the theoretically predicted $\alpha=$ $-4.0$ or $-4.4$ (and similar to other pulsars, see e.g. \citet{Geyer2017}).

We compare PSR J1823$-$3021A with the other MSP GP emitter in a GC, PSR J1824$-$2452A and two well-studied GP emitter PSR B1937$+$21 and the Crab pulsar on their energy distributions, polarisation properties and arrival time statistics. In conclusion, a broken-power law is not seen from PSR J1823$-$3021A, unlike other well-studied GP emitters. Also, PSR J1823$-$3021A is mostly unpolarised but the other GP emitters are mostly linear polarised. Regarding the waiting time distribution, PSR J1823$-$3021A shows the same exponential decay as the other GP emitters.

We compare GPs in this work with the FRBs from M81 GC. The brightest source ($S/N \approx 250$) would be detectable to distances of up to 40 kpc with MeerKAT. We note that, to be detectable at the distance of the FRB in M81 ($\approx$ 3.6 Mpc), would require a GP to be a factor of $\approx 10^{4}$ times more luminous. There is a strong periodicity in the arrival time of GPs, unlike FRBs. We found 123 GPs with multi-peak structures in one component and a GP that may exhibit quasi-periodicity in the multi-peak component. This may connect GPs to FRBs since the multi-peak component and quasi-periodicity are seen in some FRBs \citep{Pastor-Marazuela2023}. 

\section*{Acknowledgements}
We would like to express our deepest appreciation to the anonymous reviewer for the comprehensive and thoughtful review of our manuscript.
We would like to express our gratitude to Dr.\ Willem van Straten,  Dr.\ Manisha Caleb, and Andrew Jameson for kindly taking the time to answer our questions. SH is also grateful to Prof.\ Brian Schmidt, Dr.\ Robert Main, Dr.\ Kelly Gourdji, Prof.\ Ryan Shannon, Prof.\ Kenneth Freeman, Pavan Uttarkar and Pratyasha Gitika for insightful discussions. The MeerKAT telescope is operated by the South African Radio Astronomy Observatory, which is a facility of the National Research Foundation, an agency of the Department of Science and Innovation.
SH, MB and CF are supported by the Australian Research Council (ARC) Centre of Excellence (CoE) for Gravitational Wave Discovery (OzGrav) project numbers CE170100004 and CE230100016. SH is supported by the ARC CoE for All Sky Astrophysics in 3 Dimensions (ASTRO 3D) project number CE170100013.
FA acknowledges that part of the research activities described in this paper were carried out with the contribution of the NextGenerationEU funds within the National Recovery and Resilience Plan (PNRR), Mission 4 -- Education and Research, Component 2 - From Research to Business (M4C2), Investment Line 3.1 -- Strengthening and creation of Research Infrastructures, Project IR0000034 -- “STILES -- Strengthening the Italian Leadership in ELT and SKA.
This work used the OzSTAR national facility at Swinburne University of Technology. OzSTAR is funded by Swinburne University of Technology and the National Collaborative Research Infrastructure Strategy (NCRIS). Data was obtained by S Buchner in tests of the PTUSE baseband system for Open Time proposals.

\section*{Data Availability}
The data will be made publicly available on a portal under development on the OzSTAR supercomputer at Swinburne University of Technology. Early access to the raw data can be granted by emailing mbailes@swin.edu.au.



\bibliographystyle{mnras}
\bibliography{giants} 

\begin{thebibliography}{}
\makeatletter
\relax
\def\mn@urlcharsother{\let\do\@makeother \do\$\do\&\do\#\do\^\do\_\do\%\do\~}
\def\mn@doi{\begingroup\mn@urlcharsother \@ifnextchar [ {\mn@doi@} {\mn@doi@[]}}
\def\mn@doi@[#1]#2{\def\@tempa{#1}\ifx\@tempa\@empty \href {http://dx.doi.org/#2} {doi:#2}\else \href {http://dx.doi.org/#2} {#1}\fi \endgroup}
\def\mn@eprint#1#2{\mn@eprint@#1:#2::\@nil}
\def\mn@eprint@arXiv#1{\href {http://arxiv.org/abs/#1} {{\tt arXiv:#1}}}
\def\mn@eprint@dblp#1{\href {http://dblp.uni-trier.de/rec/bibtex/#1.xml} {dblp:#1}}
\def\mn@eprint@#1:#2:#3:#4\@nil{\def\@tempa {#1}\def\@tempb {#2}\def\@tempc {#3}\ifx \@tempc \@empty \let \@tempc \@tempb \let \@tempb \@tempa \fi \ifx \@tempb \@empty \def\@tempb {arXiv}\fi \@ifundefined {mn@eprint@\@tempb}{\@tempb:\@tempc}{\expandafter \expandafter \csname mn@eprint@\@tempb\endcsname \expandafter{\@tempc}}}

\bibitem[\protect\citeauthoryear{{Abbate} et~al.,}{{Abbate} et~al.}{2020}]{Abbate2020}
{Abbate} F.,  et~al., 2020, \mn@doi [\mnras] {10.1093/mnras/staa2510}, \href {https://ui.adsabs.harvard.edu/abs/2020MNRAS.498..875A} {498, 875}

\bibitem[\protect\citeauthoryear{{Abbate} et~al.,}{{Abbate} et~al.}{2022}]{Abbate2022}
{Abbate} F.,  et~al., 2022, \mn@doi [\mnras] {10.1093/mnras/stac1041}, \href {https://ui.adsabs.harvard.edu/abs/2022MNRAS.513.2292A} {513, 2292}

\bibitem[\protect\citeauthoryear{{Backer}, {Kulkarni}, {Heiles}, {Davis}  \& {Goss}}{{Backer} et~al.}{1982}]{Backer1982}
{Backer} D.~C.,  {Kulkarni} S.~R.,  {Heiles} C.,  {Davis} M.~M.,   {Goss} W.~M.,  1982, \mn@doi [\nat] {10.1038/300615a0}, \href {https://ui.adsabs.harvard.edu/abs/1982Natur.300..615B} {300, 615}

\bibitem[\protect\citeauthoryear{{Bailes} et~al.,}{{Bailes} et~al.}{2016}]{Bailes2016}
{Bailes} M.,  et~al., 2016, in MeerKAT Science: On the Pathway to the SKA. p.~11 (\mn@eprint {arXiv} {1803.07424}), \mn@doi{10.22323/1.277.0011}

\bibitem[\protect\citeauthoryear{{Bailes} et~al.,}{{Bailes} et~al.}{2020}]{Bailes2020}
{Bailes} M.,  et~al., 2020, \mn@doi [\pasa] {10.1017/pasa.2020.19}, \href {https://ui.adsabs.harvard.edu/abs/2020PASA...37...28B} {37, e028}

\bibitem[\protect\citeauthoryear{{Bhardwaj} et~al.,}{{Bhardwaj} et~al.}{2021}]{Bhardwaj2021}
{Bhardwaj} M.,  et~al., 2021, \mn@doi [\apjl] {10.3847/2041-8213/abeaa6}, \href {https://ui.adsabs.harvard.edu/abs/2021ApJ...910L..18B} {910, L18}

\bibitem[\protect\citeauthoryear{{Biggs}, {Bailes}, {Lyne}, {Goss}  \& {Fruchter}}{{Biggs} et~al.}{1994}]{Biggs1994}
{Biggs} J.~D.,  {Bailes} M.,  {Lyne} A.~G.,  {Goss} W.~M.,   {Fruchter} A.~S.,  1994, \mn@doi [\mnras] {10.1093/mnras/267.1.125}, \href {https://ui.adsabs.harvard.edu/abs/1994MNRAS.267..125B} {267, 125}

\bibitem[\protect\citeauthoryear{{Booth} \& {Jonas}}{{Booth} \& {Jonas}}{2012}]{Booth2012}
{Booth} R.~S.,  {Jonas} J.~L.,  2012, African Skies, \href {https://ui.adsabs.harvard.edu/abs/2012AfrSk..16..101B} {16, 101}

\bibitem[\protect\citeauthoryear{{Cognard}, {Shrauner}, {Taylor}  \& {Thorsett}}{{Cognard} et~al.}{1996}]{Cognard1996}
{Cognard} I.,  {Shrauner} J.~A.,  {Taylor} J.~H.,   {Thorsett} S.~E.,  1996, \mn@doi [\apjl] {10.1086/309894}, \href {https://ui.adsabs.harvard.edu/abs/1996ApJ...457L..81C} {457, L81}

\bibitem[\protect\citeauthoryear{{Cordes} \& {Lazio}}{{Cordes} \& {Lazio}}{2002}]{Cordes2002}
{Cordes} J.~M.,  {Lazio} T.~J.~W.,  2002, \mn@doi [arXiv e-prints] {10.48550/arXiv.astro-ph/0207156}, \href {https://ui.adsabs.harvard.edu/abs/2002astro.ph..7156C} {pp astro--ph/0207156}

\bibitem[\protect\citeauthoryear{{Geyer} \& {Karastergiou}}{{Geyer} \& {Karastergiou}}{2016}]{Geyer2016}
{Geyer} M.,  {Karastergiou} A.,  2016, \mn@doi [\mnras] {10.1093/mnras/stw1724}, \href {https://ui.adsabs.harvard.edu/abs/2016MNRAS.462.2587G} {462, 2587}

\bibitem[\protect\citeauthoryear{{Geyer} et~al.,}{{Geyer} et~al.}{2017}]{Geyer2017}
{Geyer} M.,  et~al., 2017, \mn@doi [\mnras] {10.1093/mnras/stx1151}, \href {https://ui.adsabs.harvard.edu/abs/2017MNRAS.470.2659G} {470, 2659}

\bibitem[\protect\citeauthoryear{{Geyer} et~al.,}{{Geyer} et~al.}{2021}]{Geyer2021}
{Geyer} M.,  et~al., 2021, \mn@doi [\mnras] {10.1093/mnras/stab1501}, \href {https://ui.adsabs.harvard.edu/abs/2021MNRAS.505.4468G} {505, 4468}

\bibitem[\protect\citeauthoryear{{Harris}}{{Harris}}{2010}]{Harris2010}
{Harris} W.~E.,  2010, \mn@doi [arXiv e-prints] {10.48550/arXiv.1012.3224}, \href {https://ui.adsabs.harvard.edu/abs/2010arXiv1012.3224H} {p. arXiv:1012.3224}

\bibitem[\protect\citeauthoryear{{Jessner} et~al.,}{{Jessner} et~al.}{2010}]{Jessner2010}
{Jessner} A.,  et~al., 2010, \mn@doi [\aap] {10.1051/0004-6361/201014806}, \href {https://ui.adsabs.harvard.edu/abs/2010A&A...524A..60J} {524, A60}

\bibitem[\protect\citeauthoryear{{Jonas}}{{Jonas}}{2009}]{Jonas2009}
{Jonas} J.~L.,  2009, \mn@doi [IEEE Proceedings] {10.1109/JPROC.2009.2020713}, \href {https://ui.adsabs.harvard.edu/abs/2009IEEEP..97.1522J} {97, 1522}

\bibitem[\protect\citeauthoryear{{Karuppusamy}, {Stappers}  \& {van Straten}}{{Karuppusamy} et~al.}{2010}]{Karuppusamy2010}
{Karuppusamy} R.,  {Stappers} B.~W.,   {van Straten} W.,  2010, \mn@doi [\aap] {10.1051/0004-6361/200913729}, \href {https://ui.adsabs.harvard.edu/abs/2010A&A...515A..36K} {515, A36}

\bibitem[\protect\citeauthoryear{{Kinkhabwala} \& {Thorsett}}{{Kinkhabwala} \& {Thorsett}}{2000}]{Kinkhabwala2000}
{Kinkhabwala} A.,  {Thorsett} S.~E.,  2000, \mn@doi [\apj] {10.1086/308844}, \href {https://ui.adsabs.harvard.edu/abs/2000ApJ...535..365K} {535, 365}

\bibitem[\protect\citeauthoryear{{Kirsten} et~al.,}{{Kirsten} et~al.}{2022}]{Kirsten2022}
{Kirsten} F.,  et~al., 2022, \mn@doi [\nat] {10.1038/s41586-021-04354-w}, \href {https://ui.adsabs.harvard.edu/abs/2022Natur.602..585K} {602, 585}

\bibitem[\protect\citeauthoryear{{Knight}}{{Knight}}{2007}]{Knight2007}
{Knight} H.~S.,  2007, \mn@doi [\mnras] {10.1111/j.1365-2966.2007.11810.x}, \href {https://ui.adsabs.harvard.edu/abs/2007MNRAS.378..723K} {378, 723}

\bibitem[\protect\citeauthoryear{{Knight}, {Bailes}, {Manchester}  \& {Ord}}{{Knight} et~al.}{2005}]{Knight2005}
{Knight} H.~S.,  {Bailes} M.,  {Manchester} R.~N.,   {Ord} S.~M.,  2005, \mn@doi [\apj] {10.1086/429533}, \href {https://ui.adsabs.harvard.edu/abs/2005ApJ...625..951K} {625, 951}

\bibitem[\protect\citeauthoryear{{Knight}, {Bailes}, {Manchester}, {Ord}  \& {Jacoby}}{{Knight} et~al.}{2006a}]{Knight2006a}
{Knight} H.~S.,  {Bailes} M.,  {Manchester} R.~N.,  {Ord} S.~M.,   {Jacoby} B.~A.,  2006a, \mn@doi [\apj] {10.1086/500292}, \href {https://ui.adsabs.harvard.edu/abs/2006ApJ...640..941K} {640, 941}

\bibitem[\protect\citeauthoryear{{Knight}, {Bailes}, {Manchester}  \& {Ord}}{{Knight} et~al.}{2006b}]{Knight2006b}
{Knight} H.~S.,  {Bailes} M.,  {Manchester} R.~N.,   {Ord} S.~M.,  2006b, \mn@doi [\apj] {10.1086/508253}, \href {https://ui.adsabs.harvard.edu/abs/2006ApJ...653..580K} {653, 580}

\bibitem[\protect\citeauthoryear{{Kramer}, {Liu}, {Desvignes}, {Karuppusamy}  \& {Stappers}}{{Kramer} et~al.}{2024}]{Kramer2024}
{Kramer} M.,  {Liu} K.,  {Desvignes} G.,  {Karuppusamy} R.,   {Stappers} B.~W.,  2024, \mn@doi [Nature Astronomy] {10.1038/s41550-023-02125-3}, \href {https://ui.adsabs.harvard.edu/abs/2024NatAs...8..230K} {8, 230}

\bibitem[\protect\citeauthoryear{{Li} \& {Pen}}{{Li} \& {Pen}}{2024}]{Li2024}
{Li} D.,  {Pen} U.-L.,  2024, \mn@doi [\mnras] {10.1093/mnras/stae1190}, \href {https://ui.adsabs.harvard.edu/abs/2024MNRAS.531.2330L} {531, 2330}

\bibitem[\protect\citeauthoryear{{Lin} \& {van Kerkwijk}}{{Lin} \& {van Kerkwijk}}{2023}]{Lin&vanKerkwijk2023}
{Lin} R.,  {van Kerkwijk} M.~H.,  2023, \mn@doi [\apj] {10.3847/1538-4357/acfa6f}, \href {https://ui.adsabs.harvard.edu/abs/2023ApJ...959..111L} {959, 111}

\bibitem[\protect\citeauthoryear{{Lin}, {van Kerkwijk}, {Main}, {Mahajan}, {Pen}  \& {Kirsten}}{{Lin} et~al.}{2023}]{Lin2023}
{Lin} R.,  {van Kerkwijk} M.~H.,  {Main} R.,  {Mahajan} N.,  {Pen} U.-L.,   {Kirsten} F.,  2023, \mn@doi [\apj] {10.3847/1538-4357/acba95}, \href {https://ui.adsabs.harvard.edu/abs/2023ApJ...945..115L} {945, 115}

\bibitem[\protect\citeauthoryear{{Lundgren}, {Cordes}, {Ulmer}, {Matz}, {Lomatch}, {Foster}  \& {Hankins}}{{Lundgren} et~al.}{1995}]{Lundgren1995}
{Lundgren} S.~C.,  {Cordes} J.~M.,  {Ulmer} M.,  {Matz} S.~M.,  {Lomatch} S.,  {Foster} R.~S.,   {Hankins} T.,  1995, \mn@doi [\apj] {10.1086/176404}, \href {https://ui.adsabs.harvard.edu/abs/1995ApJ...453..433L} {453, 433}

\bibitem[\protect\citeauthoryear{{Lynch}, {Freire}, {Ransom}  \& {Jacoby}}{{Lynch} et~al.}{2012}]{Lynch2012}
{Lynch} R.~S.,  {Freire} P. C.~C.,  {Ransom} S.~M.,   {Jacoby} B.~A.,  2012, \mn@doi [\apj] {10.1088/0004-637X/745/2/109}, \href {https://ui.adsabs.harvard.edu/abs/2012ApJ...745..109L} {745, 109}

\bibitem[\protect\citeauthoryear{{Lyutikov}, {Burzawa}  \& {Popov}}{{Lyutikov} et~al.}{2016}]{Lyutikov2016}
{Lyutikov} M.,  {Burzawa} L.,   {Popov} S.~B.,  2016, \mn@doi [\mnras] {10.1093/mnras/stw1669}, \href {https://ui.adsabs.harvard.edu/abs/2016MNRAS.462..941L} {462, 941}

\bibitem[\protect\citeauthoryear{{Mahajan} \& {van Kerkwijk}}{{Mahajan} \& {van Kerkwijk}}{2024}]{Mahajan2024}
{Mahajan} N.,  {van Kerkwijk} M.~H.,  2024, \mn@doi [\apj] {10.3847/1538-4357/ad3c35}, \href {https://ui.adsabs.harvard.edu/abs/2024ApJ...967...34M} {967, 34}

\bibitem[\protect\citeauthoryear{{Main}, {van Kerkwijk}, {Pen}, {Mahajan}  \& {Vanderlinde}}{{Main} et~al.}{2017}]{Main2017}
{Main} R.,  {van Kerkwijk} M.,  {Pen} U.-L.,  {Mahajan} N.,   {Vanderlinde} K.,  2017, \mn@doi [\apjl] {10.3847/2041-8213/aa6f03}, \href {https://ui.adsabs.harvard.edu/abs/2017ApJ...840L..15M} {840, L15}

\bibitem[\protect\citeauthoryear{{Majid}, {Naudet}, {Lowe}  \& {Kuiper}}{{Majid} et~al.}{2011}]{Majid2011}
{Majid} W.~A.,  {Naudet} C.~J.,  {Lowe} S.~T.,   {Kuiper} T. B.~H.,  2011, \mn@doi [\apj] {10.1088/0004-637X/741/1/53}, \href {https://ui.adsabs.harvard.edu/abs/2011ApJ...741...53M} {741, 53}

\bibitem[\protect\citeauthoryear{{Manchester}, {Hobbs}, {Teoh}  \& {Hobbs}}{{Manchester} et~al.}{2005}]{Manchester2005}
{Manchester} R.~N.,  {Hobbs} G.~B.,  {Teoh} A.,   {Hobbs} M.,  2005, \mn@doi [\aj] {10.1086/428488}, \href {https://ui.adsabs.harvard.edu/abs/2005AJ....129.1993M} {129, 1993}

\bibitem[\protect\citeauthoryear{{McKee} et~al.,}{{McKee} et~al.}{2019}]{McKee2019}
{McKee} J.~W.,  et~al., 2019, \mn@doi [\mnras] {10.1093/mnras/sty3058}, \href {https://ui.adsabs.harvard.edu/abs/2019MNRAS.483.4784M} {483, 4784}

\bibitem[\protect\citeauthoryear{{Nimmo} et~al.,}{{Nimmo} et~al.}{2021}]{Nimmo2021}
{Nimmo} K.,  et~al., 2021, \mn@doi [Nature Astronomy] {10.1038/s41550-021-01321-3}, \href {https://ui.adsabs.harvard.edu/abs/2021NatAs...5..594N} {5, 594}

\bibitem[\protect\citeauthoryear{{Nimmo} et~al.,}{{Nimmo} et~al.}{2023}]{Nimmo2023}
{Nimmo} K.,  et~al., 2023, \mn@doi [\mnras] {10.1093/mnras/stad269}, \href {https://ui.adsabs.harvard.edu/abs/2023MNRAS.520.2281N} {520, 2281}

\bibitem[\protect\citeauthoryear{{Oswald} et~al.,}{{Oswald} et~al.}{2021}]{Oswald2021}
{Oswald} L.~S.,  et~al., 2021, \mn@doi [\mnras] {10.1093/mnras/stab980}, \href {https://ui.adsabs.harvard.edu/abs/2021MNRAS.504.1115O} {504, 1115}

\bibitem[\protect\citeauthoryear{{Pastor-Marazuela} et~al.,}{{Pastor-Marazuela} et~al.}{2023}]{Pastor-Marazuela2023}
{Pastor-Marazuela} I.,  et~al., 2023, \mn@doi [\aap] {10.1051/0004-6361/202243339}, \href {https://ui.adsabs.harvard.edu/abs/2023A&A...678A.149P} {678, A149}

\bibitem[\protect\citeauthoryear{{Petroff} et~al.,}{{Petroff} et~al.}{2015}]{Petroff2015}
{Petroff} E.,  et~al., 2015, \mn@doi [\mnras] {10.1093/mnras/stu2419}, \href {https://ui.adsabs.harvard.edu/abs/2015MNRAS.447..246P} {447, 246}

\bibitem[\protect\citeauthoryear{{Popov} \& {Stappers}}{{Popov} \& {Stappers}}{2007}]{Popov2007}
{Popov} M.~V.,  {Stappers} B.,  2007, \mn@doi [\aap] {10.1051/0004-6361:20066589}, \href {https://ui.adsabs.harvard.edu/abs/2007A&A...470.1003P} {470, 1003}

\bibitem[\protect\citeauthoryear{{Ridolfi} et~al.,}{{Ridolfi} et~al.}{2021}]{Ridolfi2021}
{Ridolfi} A.,  et~al., 2021, \mn@doi [\mnras] {10.1093/mnras/stab790}, \href {https://ui.adsabs.harvard.edu/abs/2021MNRAS.504.1407R} {504, 1407}

\bibitem[\protect\citeauthoryear{{Romani} \& {Johnston}}{{Romani} \& {Johnston}}{2001}]{Romani2001}
{Romani} R.~W.,  {Johnston} S.,  2001, \mn@doi [\apjl] {10.1086/323415}, \href {https://ui.adsabs.harvard.edu/abs/2001ApJ...557L..93R} {557, L93}

\bibitem[\protect\citeauthoryear{{Serylak} et~al.,}{{Serylak} et~al.}{2021}]{Serylak2021}
{Serylak} M.,  et~al., 2021, \mn@doi [\mnras] {10.1093/mnras/staa2811}, \href {https://ui.adsabs.harvard.edu/abs/2021MNRAS.505.4483S} {505, 4483}

\bibitem[\protect\citeauthoryear{{Soglasnov}, {Popov}, {Bartel}, {Cannon}, {Novikov}, {Kondratiev}  \& {Altunin}}{{Soglasnov} et~al.}{2004}]{Soglasnov2004}
{Soglasnov} V.~A.,  {Popov} M.~V.,  {Bartel} N.,  {Cannon} W.,  {Novikov} A.~Y.,  {Kondratiev} V.~I.,   {Altunin} V.~I.,  2004, \mn@doi [\apj] {10.1086/424908}, \href {https://ui.adsabs.harvard.edu/abs/2004ApJ...616..439S} {616, 439}

\bibitem[\protect\citeauthoryear{Virtanen et~al.,}{Virtanen et~al.}{2020}]{scipy2020}
Virtanen P.,  et~al., 2020, \mn@doi [Nature Methods] {10.1038/s41592-019-0686-2}, \href {https://rdcu.be/b08Wh} {17, 261}

\bibitem[\protect\citeauthoryear{{Yao}, {Manchester}  \& {Wang}}{{Yao} et~al.}{2017}]{YMW2017}
{Yao} J.~M.,  {Manchester} R.~N.,   {Wang} N.,  2017, \mn@doi [\apj] {10.3847/1538-4357/835/1/29}, \href {https://ui.adsabs.harvard.edu/abs/2017ApJ...835...29Y} {835, 29}

\bibitem[\protect\citeauthoryear{{van Straten} \& {Bailes}}{{van Straten} \& {Bailes}}{2011}]{vanStraten2011}
{van Straten} W.,  {Bailes} M.,  2011, \mn@doi [\pasa] {10.1071/AS10021}, \href {https://ui.adsabs.harvard.edu/abs/2011PASA...28....1V} {28, 1}

\makeatother
\end{thebibliography}



\appendix
\section{MCMC diagnostic plots of parameter fittings} \label{appendix}
\begin{figure}
\centering
	\includegraphics[width=\linewidth]{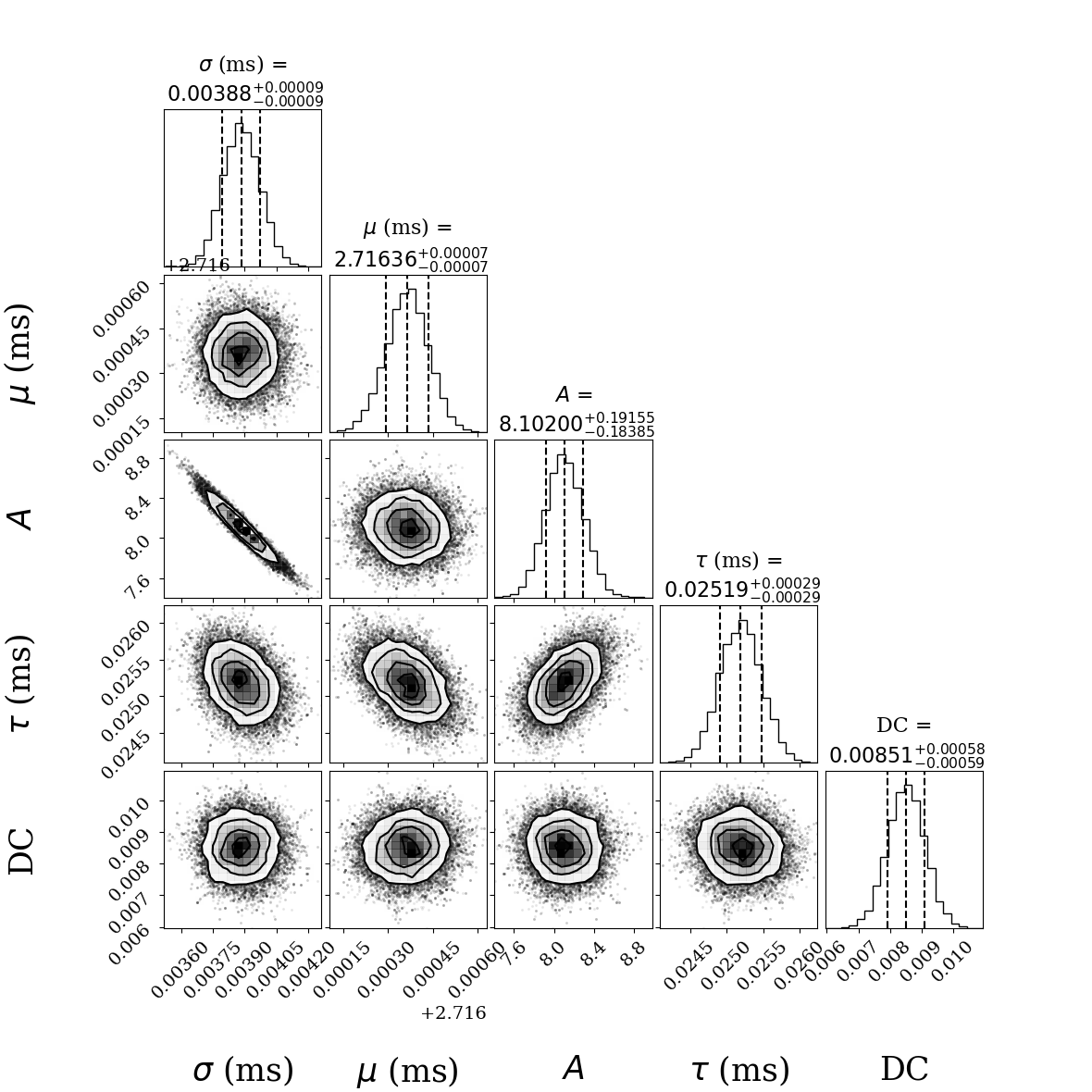}
 
    \caption{The MCMC corner plot for $\sigma$, $\mu$, $A$, $\tau$ and DC components, shown at 607.9 MHz for the brightest GP (see also Fig. \ref{scattering.png}). We used 5000 runtime steps after a burn-in phase of 2500 steps.}
    \label{mcmc_tau_freq0.pdf}
\end{figure}

\begin{figure}
\centering
	\includegraphics[width=\linewidth]{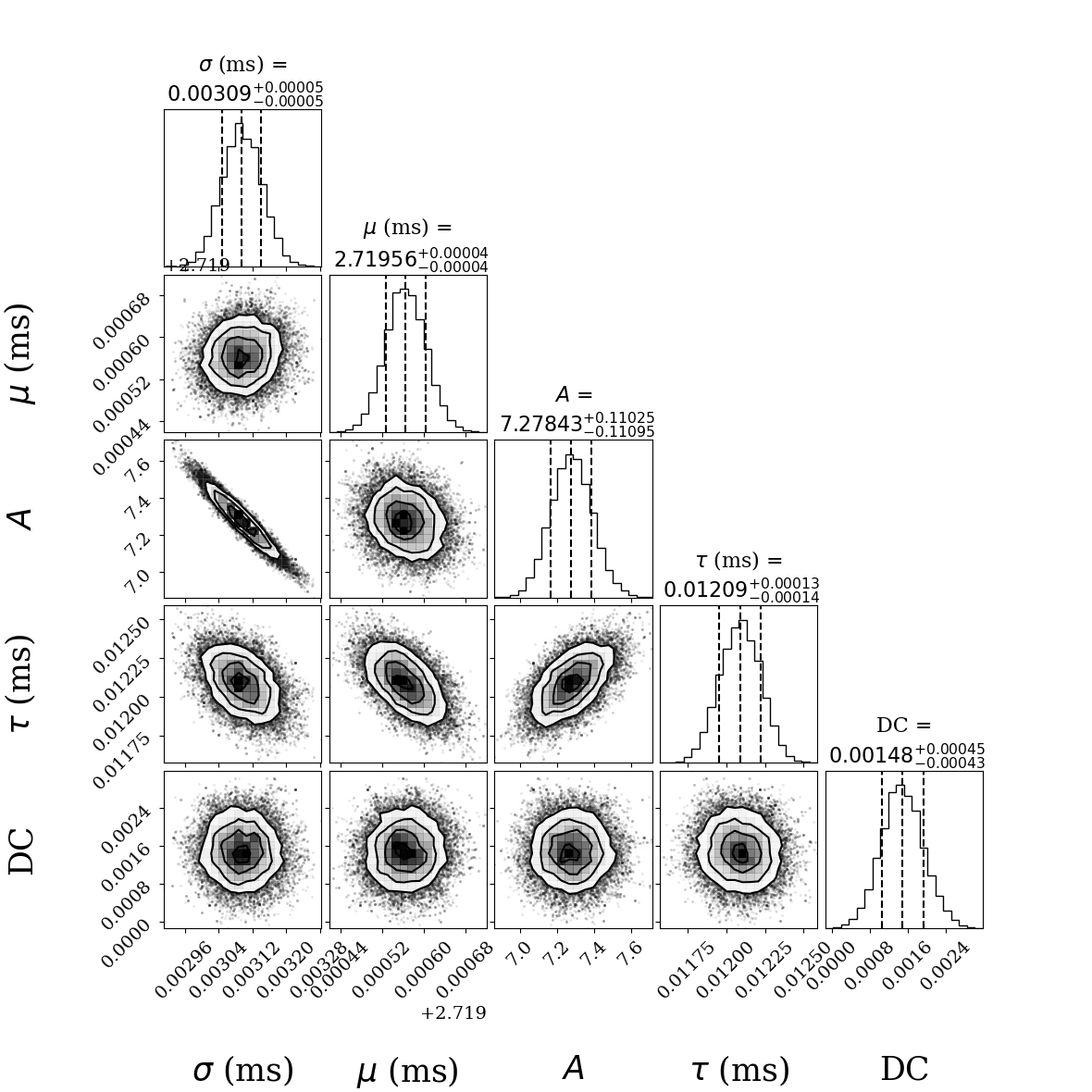}
 
    \caption{As for Fig. \ref{mcmc_tau_freq0.pdf}, but for a central frequency of 744.6 MHz.}
    \label{mcmc_tau_freq1.pdf}
\end{figure}

\begin{figure}
\centering
	\includegraphics[width=\linewidth]{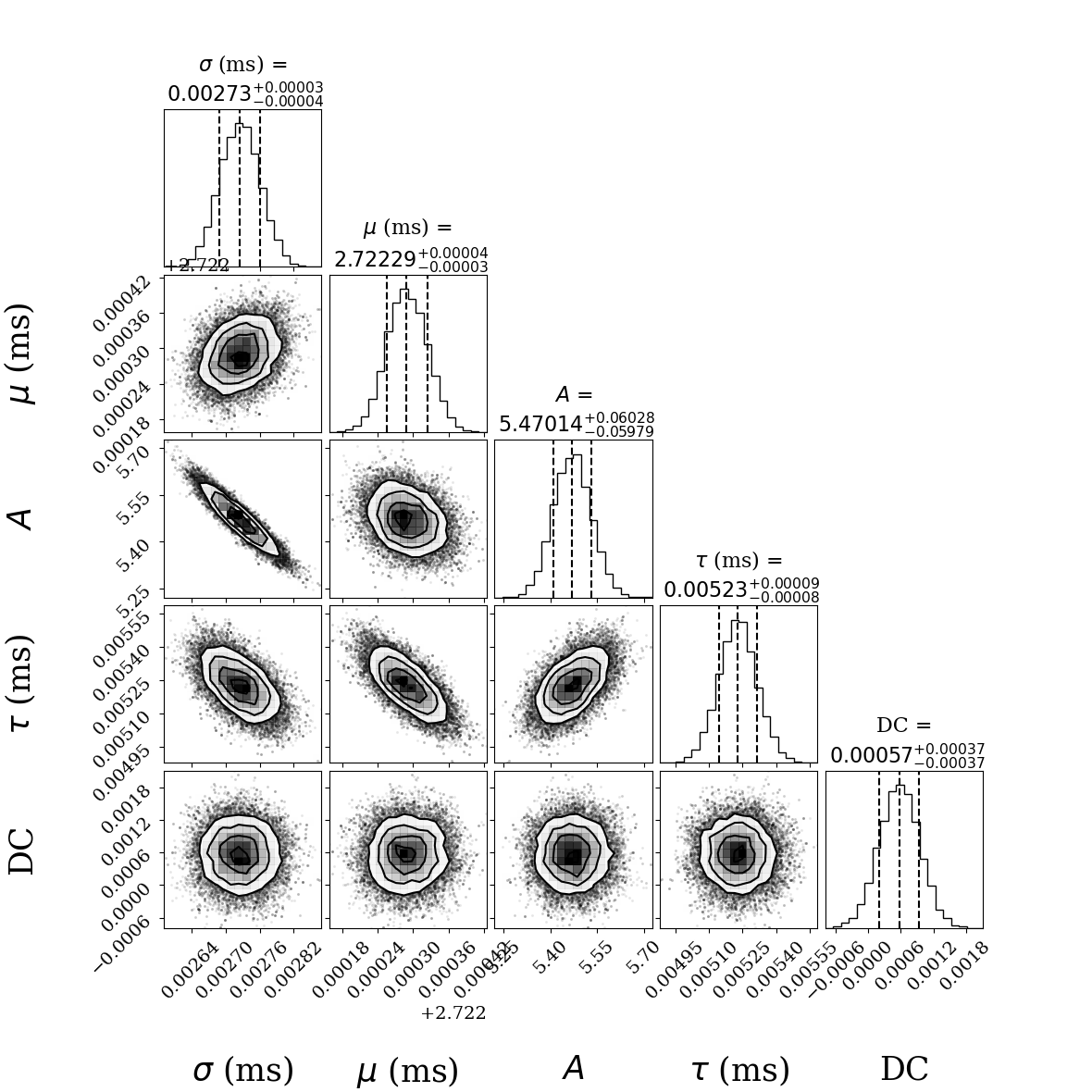}
 
    \caption{As for Fig. \ref{mcmc_tau_freq0.pdf}, but for a central frequency of  870.5 MHz}
    \label{mcmc_tau_freq2.pdf}
\end{figure}

\begin{figure}
\centering
	\includegraphics[width=\linewidth]{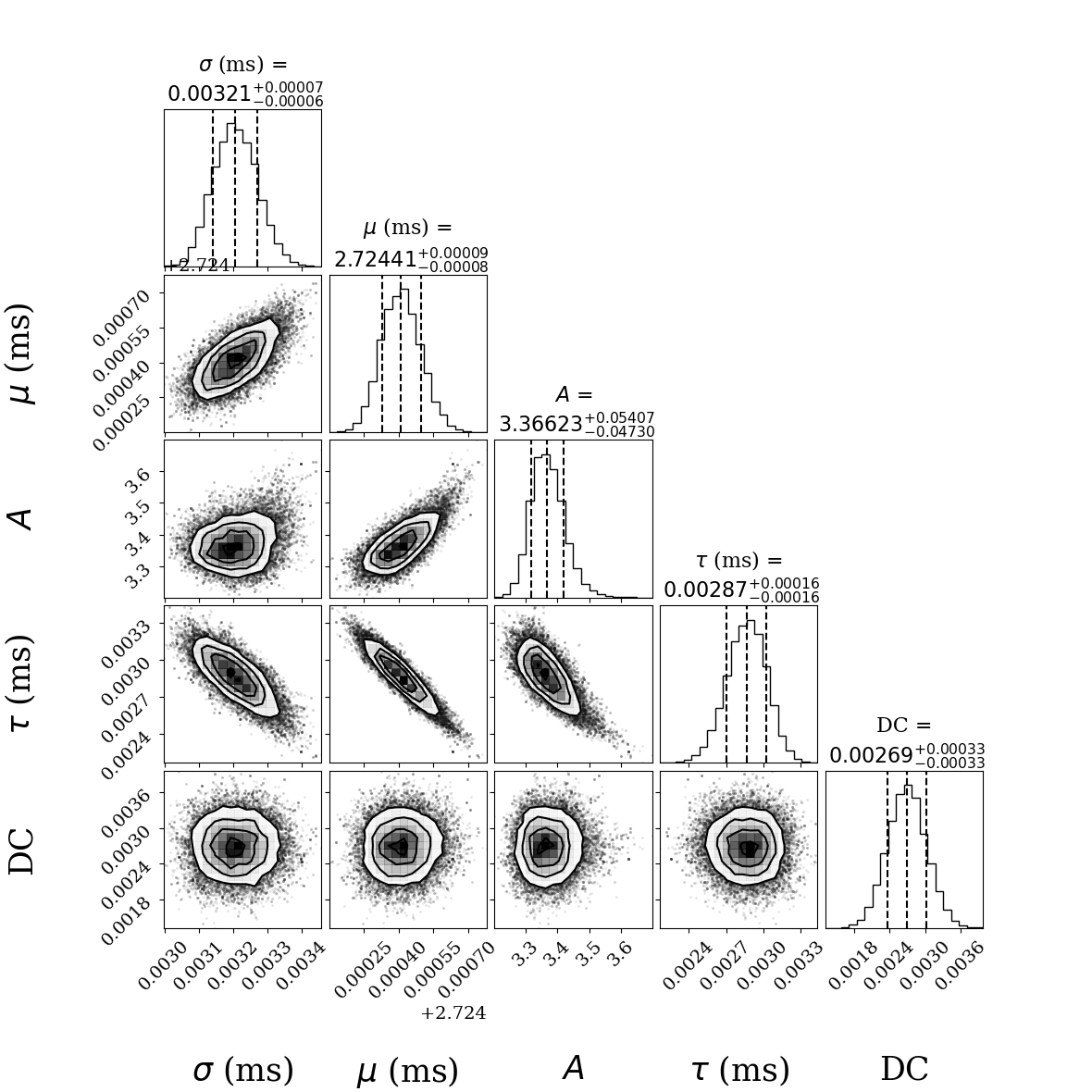}
 
    \caption{As for Fig. \ref{mcmc_tau_freq0.pdf}, but for a central frequency of 1014.2 MHz}
    \label{mcmc_tau_freq3.pdf}
\end{figure}

\begin{figure}
\centering
	\includegraphics[width=\linewidth]{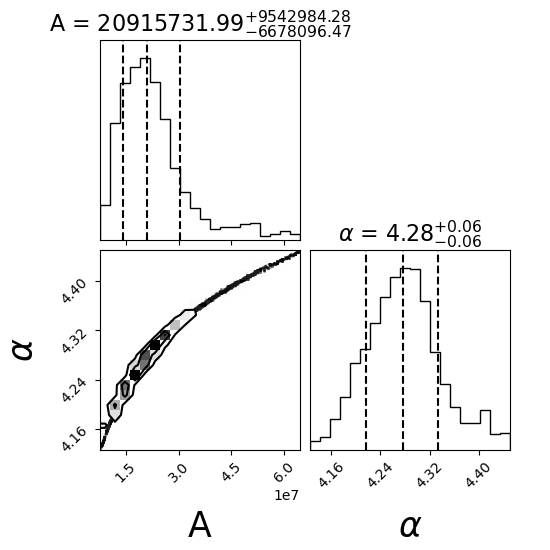}
 
    \caption{The MCMC corner plots for the $\alpha$ and A fitted for the brightest GP (see also Fig. \ref{alpha.pdf}). We used 5000 runtime steps after a burn-in phase of 2500 steps. }
    \label{mcmc_alpha.pdf}
\end{figure}

Corner plots created by the MCMC analysis in the scattering fit code \citep{Oswald2021} are presented for diagnostics. Figs. \ref{mcmc_tau_freq0.pdf}, \ref{mcmc_tau_freq1.pdf}, \ref{mcmc_tau_freq2.pdf} and \ref{mcmc_tau_freq3.pdf} show the corner plots for the parameter fits of the standard deviation ($\sigma$), mean ($\mu$), amplitude of the Gaussian ($A$), the scattering time scale ($\tau$) and any DC offset (DC) of the pulse baseline from 0 fitted for the brightest GP, shown at four frequencies, 607.9 MHz, 744.6 MHz, 870.5 MHz, and 1014.2 MHz. Fig. \ref{mcmc_alpha.pdf} shows the corner plot for the scattering index ($\alpha$) and the constant of proportionality (A) fitted for the brightest GP.

\bsp	
\label{lastpage}
\end{CJK*}
\end{document}